\documentclass{article}
\usepackage[margin=1.5in]{geometry}
\usepackage{longtable}
\usepackage{times}
\usepackage{graphicx}
\usepackage{xcolor}
\usepackage{url}
\usepackage{soul}
\usepackage{tabu}
\usepackage{algorithm}
\usepackage{algorithmic}
\usepackage{multirow}
\usepackage{amsmath}
\usepackage{comment}
\usepackage[linesnumbered,lined,boxed,commentsnumbered,algo2e]{algorithm2e}
\usepackage[misc]{ifsym}
\usepackage{amsfonts}
\usepackage{hyperref}
\SetKwRepeat{Do}{do}{while}%
\DeclareMathOperator*{\argmax}{argmax} 

\usepackage{color}
\newcommand{\blue}[1]{\textcolor[rgb]{0.00,0.00,0.00}{{#1}}}

\title{An Autonomous Performance Testing Framework using Self-Adaptive Fuzzy Reinforcement Learning}
\author{Mahshid Helali Moghadam \footnote{RISE Research Institutes of Sweden, Sweden}\and
        Mehrdad Saadatmand\footnotemark[\value{footnote}]\and
        Markus Borg\footnotemark[\value{footnote}]\and 
        Markus Bohlin\footnote{M\"alardalen University, Sweden}\and
        Bj\"orn Lisper  \footnotemark[\value{footnote}]
         }
\date{}

\begin{document}

\maketitle
\href{ }{\{mahshid.helali.moghadam, mehrdad.saadatmand, markus.borg\}@ri.se}

\href{ }{\{markus.bohlin, bjorn.lisper\}@mdh.se}
\begin{abstract}
Test automation brings the potential to reduce costs and human effort, but several aspects of software testing remain challenging to automate. One such example is automated performance testing to find performance breaking points. Current approaches to tackle automated generation of performance test cases mainly involve using source code or system model analysis or use-case based techniques. However, source code and system models might not always be available at testing time. On the other hand, if the optimal performance testing policy for the intended objective in a testing process instead could be learned by the testing system, then test automation without advanced performance models could be possible. Furthermore, the learned policy could later be reused for similar software systems under test, thus leading to higher test efficiency. We propose SaFReL, a self-adaptive fuzzy reinforcement learning-based performance testing framework. SaFReL learns the optimal policy to generate performance test cases through an initial learning phase, then reuses it during a transfer learning phase, while keeping the learning running and updating the policy in the long term. Through multiple experiments on a simulated environment, we demonstrate that our approach generates the target performance test cases for different programs more efficiently than a typical testing process, and performs adaptively without access to source code and performance models.
\end{abstract}

\section{Introduction} \label{sec::Introduction}
Quality assurance with respect to both functional and non-functional quality characteristics of software becomes crucial to the success of software products. For example, an extra one-second delay in load time of a storefront page can cause 11\% reduction in page views, and 16\% less customer satisfaction \cite{NS8}. Moreover, banking, retailing, and airline reservation systems as samples of mission-critical systems are all required to be resilient against varying conditions affecting their functional performance \cite{weyuker2000experience, brunnert2015performance, grinshpan2012solving}.  


Performance, which has been also called “efficiency” in the classification schemes of quality characteristics \cite{ISO/IEC,glinz2007non,chung2012non}, is generally referred to as how well a software system (service) accomplishes the expected functionalities. Performance requirements mainly describe time and resource bound constraints on the behavior of software, which are often expressed in terms of performance metrics such as response time, throughput, and resource utilization.


\textit{\textbf{Performance evaluation.}} Performance modeling and testing are common evaluation approaches to accomplish the associated objectives such as measurement of performance metrics, detection of functional problems emerging under certain performance conditions, and also violations of performance requirements \cite{jiang2015survey}. Performance modeling mainly involves building a model of the software system's behavior using modeling notations such as queueing networks, Markov processes, Petri nets, and simulation models \cite{cortellessa2011model,harchol2013performance,kant1992introduction}. Although models provide helpful insights into the performance behavior of the system, there are also many details of implementation and execution platform that might be ignored in the modeling \cite{denaro2004early}. Moreover, drawing a precise model expressing the performance behavior of the software under different conditions is often difficult.  
Performance testing as another family of techniques is intended to achieve the aforementioned objectives by executing the software under the actual conditions. 

Verifying robustness of the system in terms of finding performance breaking point is one of the primary purposes of performance testing. A performance breaking point refers to the status of software at  
which the system becomes unresponsive or certain performance requirements get violated.

\textit{\textbf{Research challenge.}} Performance testing to find performance breaking points remains a challenge for complex software and execution platforms. 
Testing approaches mainly raise issues of automated and efficient generation of test cases (test conditions) resulting in accomplishing the intended objective. Common approaches for generating the performance test cases such as using source code analysis \cite{zhang2012compositional}, linear programs and evolutionary algorithms on performance models \cite{zhang2002automated, gu2009search, di2007search} and UML models \cite{garousi2010genetic, garousi2008empirical, garousi2008traffic, costa2012generating, da2011generation}, using use case-based \cite{draheim2006realistic, lutteroth2008modeling}, and behavior-driven techniques \cite{schulz2019behavior,ferme2018declarative, ferme2017towards, walter2016asking} mainly rely on source code or other artifacts, which might not always be available during the testing. 

Regarding the aforementioned issues, we propose that machine learning techniques could tackle them. One category of machine learning algorithms is reinforcement learning (RL), which is mainly intended to train an agent (learner) on how to solve a problem in an environment through being rewarded or punished in a trial and error interaction with the environment. Model-free RL is a subset of RL enabling the learner to explore the environment (the behavior of the software under test (SUT) in an execution environment in our case) and learn the optimal policy, to accomplish the objective (generating performance test cases resulting in an intended performance breaking point in our case) without access to source code and a model of the system. The learner can store the learned policy and is able to replay the learned policy in future situations, which can lead to efficiency improvements.  

\textit{\textbf{Goal of the paper.}} Our research goal is represented by the following question:

\textit{How can we adaptively and efficiently generate the performance test cases resulting in the performance breaking points for different software programs without access to the underlying source code and performance models?}

Finding performance breaking point is a key purpose in robustness analysis, which is of great importance for many types of software systems, particularly in mission- and safety-critical domains \cite{fowler2009mission}. Moreover, the question above is worth exploring also in applications specifically, such as resource management (scaling, provisioning and scheduling) for cloud services \cite{jennings2015resource}, performance prediction \cite{venkataraman2016ernest, kolesnikov2019tradeoffs}, and performance analysis of software services in other areas 
\cite{morabito2017virtualization, babovic2016web}.   

\textit{\textbf{Contribution.}} In this paper, we present the design and experimental evaluation of a self-adaptive fuzzy reinforcement learning-based (SaFReL) performance testing framework. It is intended to efficiently and adaptively generate the (platform-based) performance test conditions leading to the performance breaking point for different software programs with different performance sensitivity to resources (e.g., CPU-, memory-, and disk-intensive programs) without access to source code and performance models. \blue{An early-stage general formulation of the idea
of using RL particularly in performance testing was introduced in our prior work \cite{moghadam2019machine}. The initial formulation introduces a single smart tester agent that uses RL (simple Q-learning) in a two-phase learning together with an initial architecture in the abstract.    
This paper extends the initial abstract formulation of the RL-assisted performance testing \cite{moghadam2019machine}. It uses an elaborate learning technique originally inspired by the conference paper by  \cite{ibidunmoye2017adaptive}, which presents an adaptive performance (response time) control approach for cloud services using cooperative fuzzy multi-agent reinforcement learning.
However, regarding the distinguishing learning details, the proposed RL-assisted performance testing framework is based on a single smart agent, involves two distinct phases of learning, and benefits a particular adaptive learning strategy which plays an important role in the functionality of the agent.
The proposed smart performance testing framework is intended to conduct performance testing to meet a testing objective that is finding an intended performance breaking point. The proposed framework, SaFReL, is a two-phase RL-assisted performance testing agent that is able to learn the efficient generation of performance test cases to meet the testing objective and more importantly replay the learned policy in further similar testing situations.} 

\blue{SaFReL assumes two phases of learning: initial and transfer learning. In the initial learning phase, it learns the optimal policy to generate the target performance test cases initially upon observing the behavior of the first SUT. Afterward in the transfer learning, it reuses the learned policy for the SUTs with a performance sensitivity analogous to already observed ones while still keeping the learning running in the long term.
The learning mechanism uses Q-learning augmented by fuzzy logic in one part of the learning to deal with the issue of uncertainty in defining discrete categories over continuous values as used by \cite{ibidunmoye2017adaptive}. The single light-weight RL tester agent has the capability of transfer learning and reusing knowledge in similar situations. It benefits an adaptive action selection strategy that adapts the learning to various testing situations and subsequently makes the agent able to act efficiently on various SUTs.} 

We demonstrate that SaFReL works adaptively and efficiently on different sets of SUTs, which are either homogeneous or heterogeneous in terms of their performance sensitivity.
Our experiments are based on simulating the performance behavior of 50 instances of 12 well-known programs as the SUTs. Those instances are characterized by various initial amounts of granted resources and different values of response time requirements. We use two evaluation criteria, namely efficiency and adaptivity, to evaluate our approach. 
We investigate the efficiency of the approach in generating the test cases that result in reaching the intended performance breaking point and also the behavioral sensitivity of the approach to the learning parameters.  
\blue{In particular, SaFReL reaches the intended objective more efficiently compared to a typical stress testing technique, which generates the performance test cases based on changing the conditions, e.g., decreasing the availability of resources, by certain steps in an exploratory way.}  
SaFReL leads to reduced cost (in terms of computation time) for performance test case generation by reusing the learned policy upon the SUTs with similar performance sensitivity. Moreover, it adapts its operational strategy to various SUTs with different performance sensitivity effectively while preserving efficiency. To summarize, our contributions in this paper are:
\begin{itemize}
\item A smart performance testing framework (agent) that learns the optimal policy (way) to generate the performance test cases meeting the testing objective without access to source code and models, and reuses the learned policy in further testing cases. 
It uses fuzzy RL and an adaptive action selection strategy for the generation of test cases, and implements two phases of learning:
\begin{itemize}
\item Initial learning during which the agent learns the optimal policy for the first time,
\item Transfer learning during which the agent replays the learned policy in similar cases while keeping the learning running in the long term. 
\end{itemize}
\item A two-fold experimental evaluation involving performance (efficiency and adaptivity) and sensitivity analysis of the approach.\\
\blue{The evaluation is carried out based on simulating the performance behavior of various SUTs. We use a performance simulation module instead of actually executing SUTs. The main function of the performance simulation module is estimating the performance behavior of SUTs in terms of their response time.}

\end{itemize}

\textit{\textbf{Structure of the paper.}} The rest of the paper is organized as follows: Section \ref{sec::Motivation and Background} discusses the background concepts and motivations for the proposed self-adaptive learning-based approach. Section \ref{sec::System Architecture} presents an overview of the architecture of the proposed testing framework, while the technical details of the constituent parts are described in Sections \ref{sec::Fuzzy State Detection} and \ref{sec::Adaptive Action Selection and Reward}. In Section \ref{sec::Stress Testing using Self-Adaptive}, we explain the functions of the learning phases. Section \ref{sec::Evaluation} reports on the experimental evaluation involving the experiment's setup, and the results of the experimentation. Section \ref{sec::Discussion} discusses the results, the lessons learned during the experimentation, and also the threats to the validity of the results. Section \ref{sec::Related Work} provides a review on the related work, and finally Section \ref{sec::Conclusion} concludes the paper and discusses some future directions.

\section{Motivation and Background}\label{sec::Motivation and Background}
Performance analysis, realized through modeling or testing, is important for performance-critical software systems in various domains. 
Anomalies in performance behavior of a software system or violations of performance requirements are generally consequences of the emergence of performance bottlenecks at the system or platform levels \cite{ibidunmoye2015performance, chandola2009anomaly}. A performance bottleneck is a system or resource component limiting the performance of the system and hinders the system from acting as required \cite{gregg2013systems}. 
The behavior of a bottleneck component is due to some limitations associated with the component such as saturation and contention. A system or resource component saturation happens upon full utilization of its capacity or when the utilization exceeds a usage threshold \cite{gregg2013systems}. Capacity expresses the maximum available processing power, service (giving) rate, or storage size. Contention occurs when multiple processes contend for accessing a limited number of shared components including resource components like CPU cycles, memory, and disk or software (application) components.

There are various application-, platform- and workload-based causes for the emergence of performance bottlenecks \cite{ibidunmoye2015performance}. 
Application-based causes represent issues such as defects in the source code or system architecture faults. Platform-based causes characterize the issues related to hardware resources, operating system, and execution platform. High deviations from the expected workload intensity and similar issues such as workload burstiness are denoted by workload-based causes. 

On the other hand, detecting violations of performance requirements and finding performance breaking points are challenging, particularly for complex software systems. To address these challenges, we need to find how to provide critical execution conditions that make the performance bottlenecks emerge. The focus of performance testing in our case is to assess the robustness of the system and find the performance breaking point.

The effects of the internal causes (application/architecture-based ones) could vary, e.g., due to continuous changes and updates of the software during Continuous Integration/Continuous Delivery (CI/CD), and even vary upon different execution platforms and under different workload conditions. Therefore, the complexity of SUT and a variety of affecting factors make it hard to build a precise performance model expressing the effects of all types of factors at play. This is a major barrier motivating the use of model-free learning-based approaches like model-free RL in which the optimal policy for accomplishing the objective could be learned indirectly through interaction with the environment (SUT and the execution platform). In this problem statement,
the testing system learns the optimal policy to achieve the target that is finding an intended performance breaking point, for different types of software without access to a model of the environment. The testing system explores the behavior of the SUT through varying the platform-based (and workload-based in future work) test conditions, stores the learned policy and is able to later reuse the learned policy in similar situations, i.e., other SUTs with similar performance sensitivity to resource restriction. This is the feature of the proposed learning approach that is supposed to lead to a considerable reduction in the testing system's effort, and subsequently saving computation time.

Regarding the aforementioned challenges and strong points of the model-free learning-based approach, we hypothesize that in a CI/CD process based on agile software development, performance engineers and testers can save time and resources by using SaFReL for performance (stress) testing of various releases or variants. SaFReL provides an agile efficient performance test case generation technique (See Section \ref{sec::Evaluation} and Section \ref{sec::Discussion} for efficiency evaluation) while eliminating the need for source code or system model analysis.     

\subsection{Reinforcement Learning}\label{sec::Reinforcement Learning}
\blue{Reinforcement learning (RL) \cite{sutton2018reinforcement} is a fundamental category of machine learning algorithms generally intended to find the optimal behavior (way) in decision-making problems. RL is an interactive learning paradigm that is different from the common supervised and unsupervised machine learning algorithms and has been frequently applied to building many self-adaptive smart systems. It involves continuous interaction between the agent (learner) and the environment that is controlled. At each step of the interaction, the agent observes (senses) the \textit{state} of the environment, takes a possible \textit{action} and receives a reinforcement signal as a scalar \textit{reward} from the environment that shows the effectiveness of the applied action to guide the agent towards accomplishing the intended objective. There is no supervisor in RL, and the agent just receives a reward signal. RL basically involves a sequential decision-making process. The RL agent goes through the environment, decides how to behave at each step, and based on optimizing the long-term received reward, learns the optimal way of decision making. }

\begin{figure}[ht]
\centering
\includegraphics[width=.40\textwidth, height=4cm]{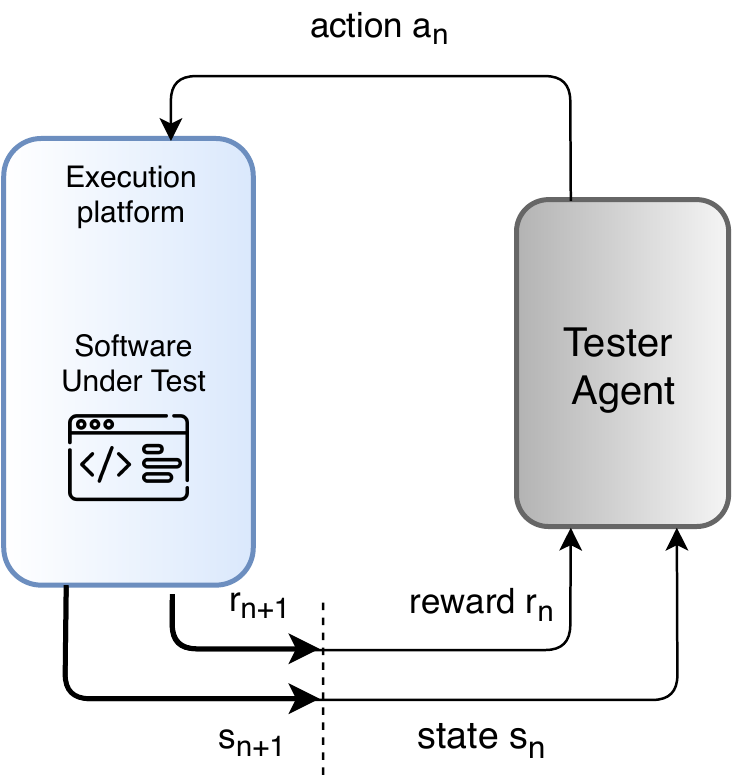}
\caption{\blue{Interaction between agent and SUT in RL} }
\label{fig: RL}
\end{figure}

\blue{The agent actually decides between actions based on the history of its observations. However, considering the whole history of observations is not efficient, therefore, \textit{state} should be formulated as a concise summary of the history including all the required information. Keeping in mind this issue, a related helpful concept to formulate the state as a summary function is the \textit{Markov state}. The states of the environment are Markov by definition. Then, when the environment is fully observable to the agent, the states that the agent observes and uses for making decisions, are Markov too. The environment in our case is the SUT and the execution platform.
The state is modeled in terms of response time and resource utilization improvement. The actions are some operations for modifying/adjusting the available capacity of resources and the objective of the agent is finding an intended performance breaking point. Figure \ref{fig: RL} shows the interaction between the agent and the environment \blue{that is the composition of SUT and execution platform in our case}.}

\blue{There are three main elements in an RL agent: policy, value function, and model. The Policy is the behavior function describing what actions the agent takes in a certain state. Value function indicates how good each state and/or action is, in terms of the amount of reward expected upon taking a particular action given a particular state. Finally, the model is the agent's view of the environment and describes what the environment does next, e.g., shows the state transitions of the environment.} 

\blue{Model-free RL algorithms are special types of RL that are not intended to build or learn a model of the environment. Instead, they learn the optimal behavior to achieve the intended objective through multiple experiences of interaction with the environment.} 
\blue{Temporal Difference (TD) \cite{sutton2018reinforcement} is one of the main types of model-free RL, which is able to learn from the incomplete episodes of the interaction with the environment. Q-learning as a model-free TD learns the optimal policy through learning the optimal value function, i.e., Q-values. It uses an action selection strategy based on a combination of trying out the available actions, namely exploration, and relying on the previously achieved experience to select the highly-valued actions, namely exploitation. It is off-policy, which means that the agent learns the optimal policy regardless of how the agent explores the environment. After learning the optimal policy, in the transfer learning phase, the agent is able to replay the learned policy while keeping the learning running, which implies occasionally exploring the action space and trying out different actions.}


\section{Architecture}\label{sec::System Architecture}
\begin{figure}[ht]
\centering
\includegraphics[width=.99\textwidth]{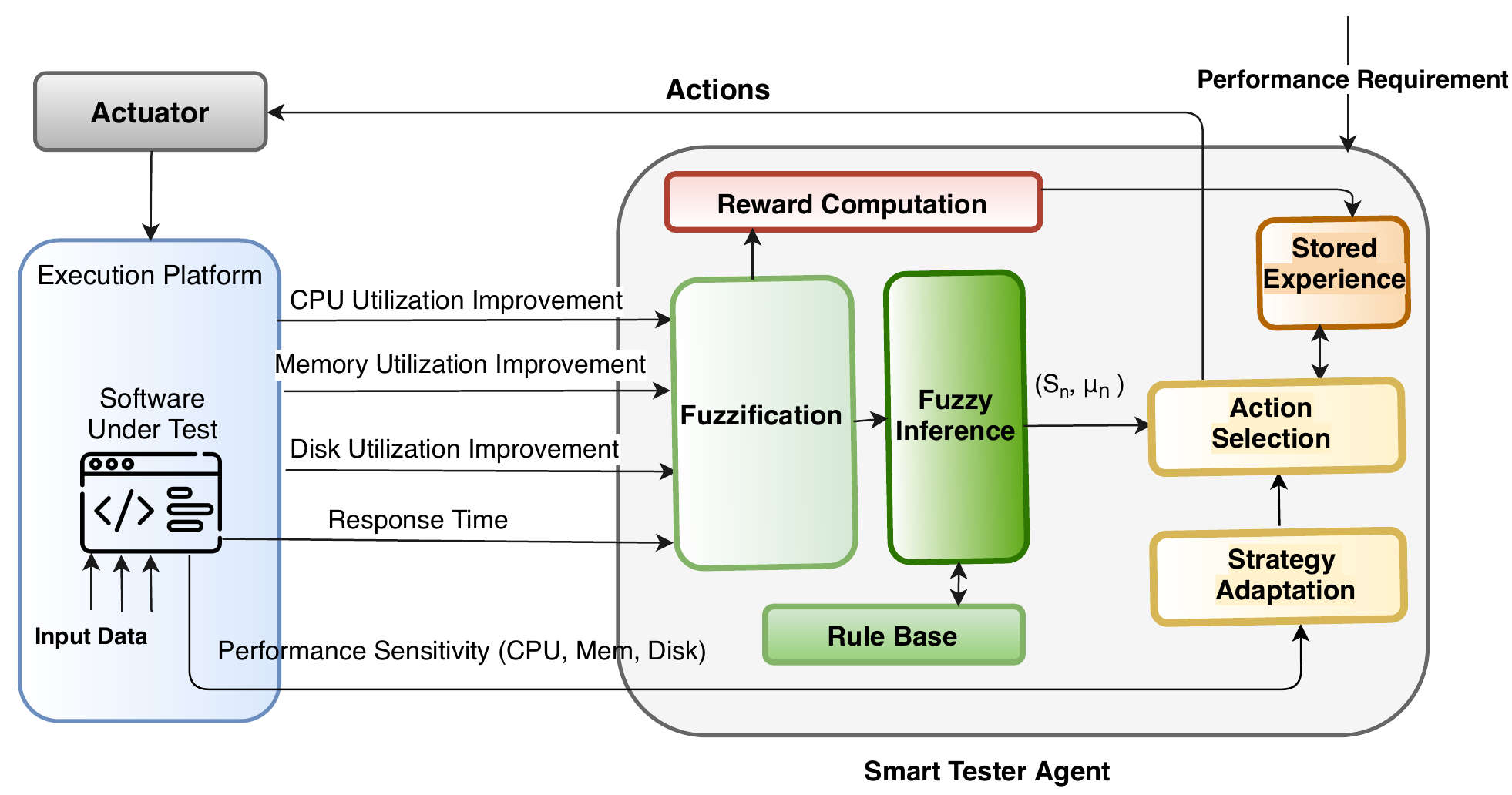}
\caption{\blue{SaFReL architecture} }
\label{fig: SaFReL architecture}
\end{figure}

This section provides an overview of the architecture of the proposed smart performance testing framework, SaFReL (see Figure \ref{fig: SaFReL architecture}). The entire interaction of the smart framework with each SUT, as a learning episode, consists of a number of learning trials. The steps of learning in each trial and the components involved in each step are described as follows: 
\paragraph{\textit{1. Fuzzy State Detection.}} The fuzzification, fuzzy inference, and rule base components in Figure \ref{fig: SaFReL architecture} are involved in the state detection. The agent uses the values of four quality metrics, 1) response time, and utilization improvements of 2) CPU, 3) memory, and 4) disk, to identify the state of the environment. In other words, the \textit{state} expresses the status of the environment relative to the testing target. In our case, these quality metrics are used to model (represent) the state space of the environment. An ordinary approach for state modeling in RL problems is dividing the state space into multiple mutually exclusive discrete sets. Each set represents a discrete state. At each time, the environment must be at one distinct state. The relevant challenges of such crisp categorization or defining discrete states, include knowing how much a value is suitable to be a threshold for categories of a metric, and how we can treat the boundary values between categories. Instead of crisp discrete states, using fuzzy logic and defining fuzzy states can help address these challenges. We use fuzzy classification as a soft labeling technique for presenting the values of the metrics used for modeling the state of the environment. Then, using a fuzzy inference engine and fuzzy rule base, the agent detects the fuzzy state of the environment. More details about the fuzzy state detection of the agent are presented in Section \ref{sec::Fuzzy State Detection}.
\paragraph{\textit{2. Action Selection and Strategy Adaptation.}} After detecting the fuzzy state of the SUT, the agent takes an action. The actions are operations modifying the factors affecting the performance, i.e., the available resource capacity, in the current prototype. The agent selects the action according to an \textit{action selection strategy} that it follows. The action selection strategy determines to what extent the agent should explore and try out the available actions, and to what extent it should rely on the learned policy and select a high-value action that has been tried and assessed before. The role of this strategy is guiding the action selection of the agent throughout the learning and is of importance for the efficiency of the learning. In order to obtain the desired efficiency, a proper trade-off between the exploration of the state action space and exploitation of the previously learned policy is critical.

In our proposed framework, the smart agent is augmented by a \textit{strategy adaptation} characteristic, as a meta-learning feature responsible for dynamically adapting the degree of exploration and exploitation in various situations. This feature makes SaFReL able to detect where it should rely on the previously learned policy and where it should make a change in the strategy to update its policy and adapt to new situations. New situations mean acting on new SUTs that are different from the previously observed ones in terms of performance sensitivity to resources.

\blue{Software programs have different levels of sensitivity to resources. SUTs with different performance sensitivity to resources, e.g., CPU-intensive, memory-intensive, or disk-intensive SUTs, will react to changes in resource availability differently. Therefore, when the agent observes a SUT that is different from the previously observed ones in terms of performance sensitivity, the strategy adaptation tries to guide the agent towards doing more exploration than exploitation. A performance sensitivity indicator showing the sensitivity of SUT to the resources (i.e., being CPU-intensive, memory-intensive or disk-intensive) is an input to the strategy adaptation mechanism (see Figure \ref{fig: SaFReL architecture}).}

The components corresponding to the action selection, the stored experience (learned policy), and the strategy adaptation are shown as yellow components in Figure \ref{fig: SaFReL architecture}. More details about the set of actions and the mechanism of strategy adaptation are described in Section \ref{sec::Adaptive Action Selection and Reward}.
\paragraph{\textit{3. Reward Computation.}} After taking the selected action, the agent receives a reward signal indicating the effectiveness of the applied action to approach the intended performance breaking point. The reward computation component (red block) in Figure \ref{fig: SaFReL architecture} calculates the received reward (see Section \ref{sec::Adaptive Action Selection and Reward}) for the taken actions.

\section{Fuzzy State Detection} \label{sec::Fuzzy State Detection}
The state space of the environment in our learning problem is modeled by the quality measurements, CPU, memory, and disk resource utilization improvement and response time of the SUT, which is shown in Figure \ref{fig: Fuzzy representation of quality measurements}.
\blue{The learning approach works based on detecting (discrete) states of the system. These states could be typically defined based on classifying the continuous values of the quality measurements that were mentioned above. On the other hand, defining such crisp boundaries on a number of continuous domains is an issue that might involve many uncertainties. 
In order to address this issue and preserve the desired precision of the model, fuzzy classification and reasoning is used to specify the states of the system. Therefore, the states of the environment are defined in terms of some fuzzy states and the environment can be in one or more fuzzy states at the same time with different degrees of certainty. The agent detects the state of the system using a fuzzy inference engine and a rule base \cite{kuncheva2008fuzzy, Fuzzyinference} (Figure \ref{fig: SaFReL architecture}). In summary, the step of state detection is done based on making fuzzy inference about the state of the system. The fuzzy state detection consists of three main parts: normalization of the input values (quality measurements), fuzzification of the measurements, and the fuzzy inference to identify the state of the environment. The details of these parts together with the fuzzy rules, fuzzy operators, and the implication method that are used, are described in Section \ref{sec::Fuzzy State Space Modelling}.}  

\subsection{State Modeling and Fuzzy Inference} \label{sec::Fuzzy State Space Modelling}
\textbf{\textit{Normalization.}} As described in the previous section, a set of quality measurements, CPU, memory, and disk utilization improvements and response time of the SUT, represent the state of the environment. The values of these measurements are not bounded, then for simplifying the inference and also the exploration of the state space, we normalize the values of these parameters to the interval [0, 1] using the following functions:

\begin{equation}\label{eq:1}
{RT_n}=\frac{2}{\pi}\tan^{-1}(\frac{RT_n^\prime}{RT^q})
\end{equation}

\begin{align}\label{eq:2}
{CUI}_n&=\frac{1}{CUI_n^\prime} & MUI_n&=\frac{1}{MUI_n^\prime} & DUI_n&=\frac{1}{DUI_n^\prime}
\end{align}

where $RT_n^\prime$, \(CUI_n^\prime\), \(MUI_n^\prime\), and \(DUI_n^\prime\) are the measured values of the response time, CPU, memory and disk utilization improvements at time step \(n\) respectively and \(RT^q\) is the response time requirement. \(CUI_n^\prime\) as the CPU utilization improvement is the ratio between the CPU utilization at time step \(n\) and its initial value (at the start of learning), that is, \({CUI_n^\prime}=\frac{CU_n}{CU^i}\). Likewise, those are, \({MUI_n^\prime}=\frac{MU_n}{MU^i}\) and \({DUI_n^\prime}=\frac{DU_n}{DU^i}\). Using the normalization function in Eq. \ref{eq:1}, when \({RT_n^\prime}=RT^q\) the normalized value of the response time, \(RT_n\) is 0.5, and for \({RT_n^\prime}> RT^q\) the normalized values will be toward 1 and for \({RT_n^\prime}< RT^q\) the normalized values will be toward 0. A tuple as \((CUI_n, MUI_n, DUI_n, RT_n)\) consisting of the normalized values of quality measurements is the input to the fuzzy state detection.

\begin{figure}[ht]
\centering
\includegraphics[width=.9\textwidth]{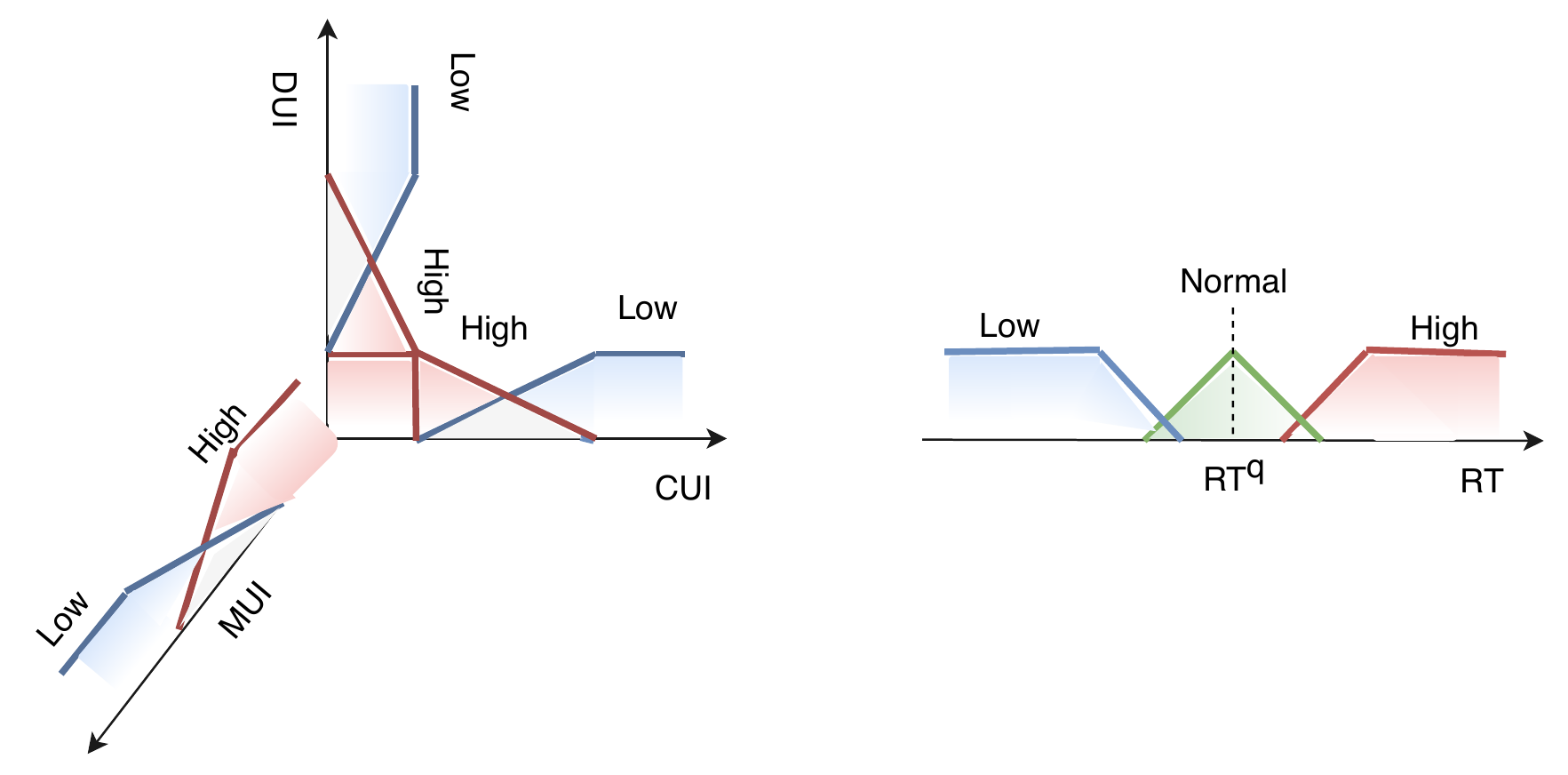}
\caption{Fuzzy representation of quality measurements}
\label{fig: Fuzzy representation of quality measurements}
\end{figure}

\textbf{\textit{Fuzzification.}} Input fuzzification involves defining fuzzy sets and corresponding membership functions over the values of the quality measurements. A membership function is characterized by a linguistic term. A fuzzy set $L$ is defined as $L=\{(x, \mu_L(x))|\ 0<x\textrm{,}\quad x\in \mathbb{R}\}$ where a membership function $\mu_L(x)$ defines membership degrees of the values as $\mu_L:x\rightarrow[0,1]$. Figure \ref{fig: Fuzzy representation of quality measurements} shows the membership functions defined over the value domains of quality measurements. As shown in Figure \ref{fig: Fuzzy representation of quality measurements}, trapezoidal membership functions are used for \textit{High} and \textit{Low} fuzzy sets and a triangular counterpart for the \textit{Normal} fuzzy set on the response time. In Figure \ref{fig: Fuzzy representation of quality measurements}, where \(RT^q\) is the requirement, a normal (medium) fuzzy set over the values of response time implies a small range around the requirement value as normal response time values. Moreover, in this case the ranges of membership functions were selected empirically and could be updated based on the requirements.

\textbf{\textit{Fuzzy Inference.}} After input fuzzification, inferring the possible states that the environment assumes is directed by the fuzzy rules that have formed based on the domain knowledge.

\textit{Fuzzy Rules.}  A fuzzy rule, as shown in Eq. \ref{eq:fuzzy rule}, consists of two parts: antecedent and consequent. The former is a combination of linguistic terms of the input normalized quality measurements and the consequent is a fuzzy set with a membership function showing to what extent the environment is in the associated state.
\begin{equation} \label{eq:fuzzy rule}
\begin{split}
\textrm{Rule 1: } & \textrm{If CUI is High AND MUI is High AND DUI is Low AND}\\
&\textrm{RT is Normal, then State is HHLN}. 
\end{split}
\end{equation}
\blue{$Rule~1$ is a sample of the fuzzy rules in the rule base. The rest of the rules are defined similarly based on the fuzzy sets defined over the values of the quality measurements and the combinations of them. Based on the number of fuzzy sets, namely two fuzzy sets, \textit{High} and \textit{Low}, over the value range of each resource utilization improvement and three sets, \textit{High}, \textit{Normal}, and \textit{Low},  over the value range of the response time, we define 24 rules in our rule base to define the fuzzy states of the environment.} 

\textit{Fuzzy Operators.} When the antecedents of the rules are made of multiple linguistic terms, which are associated to fuzzy sets, e.g., "High, High, Low and Normal", then fuzzy operators are applied to the antecedent to obtain one number showing the support or activation degree of the rule. Two well-known methods for the fuzzy \(AND\) operator are \(minimum (min)\) and \(product (prod)\). In our case, we use method \(min\) for the fuzzy \(AND\) operation. It shows that given a set of input parameters \(A\), the degree of support for rule \(Ri\) is given as \(\tau_{Ri}=\min\limits_j \mu_L(a_j)\) where \(a_j\) is an input parameter in A and L is its associated fuzzy set in the rule Ri.

\textit{Implication Method.} After obtaining the membership degree for the antecedent, the membership function of the consequent is reshaped using an implication method. There are also two well-known methods for implication process, \(minimum (min)\) and \(product (prod)\), which truncate and scale the membership function of the output fuzzy set respectively. The membership degree of the antecedent is given as input to the implication method. We use method \(min\) as the implication method in our case.

\begin{figure}[ht]
\centering
\includegraphics[width=.75\textwidth, height=6cm]{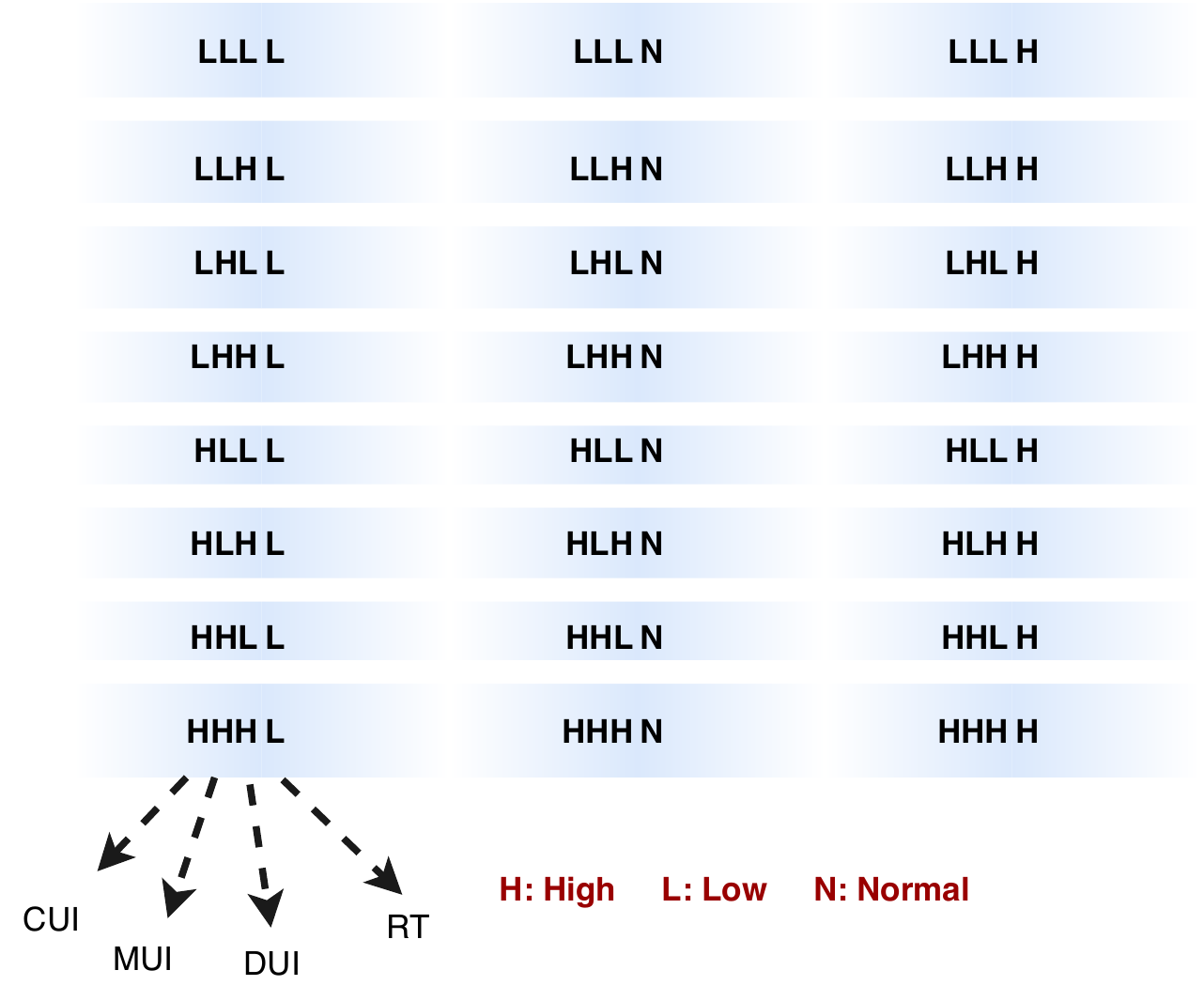}
\caption{Fuzzy states of the environment}
\label{fig: Fuzzy states of SUT}
\end{figure}

\blue{Finally, 
the most effective rule, 
the one with the maximum support degree, is selected to determine the final fuzzy state of the environment \({(S_n,\mu_n)}\). In summary, the fuzzy state with the highest likelihood is considered as the state of the system.} Figure \ref{fig: Fuzzy states of SUT} shows a representation of the fuzzy states. Each of them represents one state based on the fuzzy values (linguistic terms) assigned to quality measurements (CPU, memory and disk utilization improvement, and response time). Regarding the presentation of fuzzy states, L, H, and N stand for Low, High, and Normal terms respectively.

\section{Adaptive Action Selection and Reward Computation} \label{sec::Adaptive Action Selection and Reward}
\textit{\textbf{Actions.}} In SaFReL, the actions are the operations changing the platform-based factors affecting the performance, i.e., the available resources such as computation (CPU), memory and disk capacity. In the current prototype, the set of actions contains operations reducing the available resource capacity with finely tuned steps, which are as follows:

\begin{equation} \label{eq:AC}
\begin{split}
AC_n=&\{\textrm{no action}\}\ \cup\ \{(CPU_n-y)\ |\ y \in CDF\}\ \cup\ \{(Mem_n-k)\ |\ k \in MDF_n \}\\ 
&\cup\ \{(Disk_n-k)\ |\ k \in MDF_n\}
\end{split}
\end{equation}
\begin{equation} \label{eq:CDF}
CDF=\{\frac{1}{4},\frac{2}{4},\frac{3}{4},1\}
\end{equation}
\begin{equation} \label{eq:MDF}
MDF_n=\{(x\times\frac{Mem(Disk)_n}{4})\ |\  x \in \{\frac{1}{4},\frac{2}{4},\frac{3}{4},1\}\}
\end{equation}
where \(AC_n\), \(CPU_n\), \(Mem_n\) and \(Disk_n\) represent the set of actions, the current available computation (CPU), memory and disk capacity at time step n respectively. The list of actions is as shown in Table \ref{table: Actions in SaFReL}.\\
\begin{table}[b]
\begin{center}
\caption{Actions in SaFReL}
\begin{tabular}{ |p{6 cm}|p{4cm}|}
 \hline
 \multicolumn{2}{|c|}{\textbf{\textit{Actions}}} \\
\hline
 \textbf{\textit{Operation}} & \textbf{\textit{Decrease}} \\
 \hline
 Reducing memory / disk capacity & by a factor in \(MDF_n\) \\
 \hline
 Reducing computation (CPU) capacity & by a factor in CDF\\
 \hline
 No action & - \\
 \hline
\end{tabular}
\label{table: Actions in SaFReL}
\end{center}
\end{table}

\textit{\textbf{Strategy Adaptation.}} The agent can use different strategies for selecting the actions. $\varepsilon$-greedy with different $\varepsilon$-values and Softmax are well-known methods for action selection in RL algorithms. They are intended to provide a right trade-off between exploration of the state action space and exploitation of the learned policy. In SaFReL, we use $\varepsilon$-greedy as the action selection strategy and the proposed strategy adaptation feature acts as a simple meta-learning algorithm intended to make changes to the $\varepsilon$ value dynamically to make the action selection strategy well-adapted to new situations (new SUTs). Upon observing a SUT instance with a performance sensitivity different from the already observed ones, it adjusts the value of the parameter $\varepsilon$ to direct the agent toward more exploration (setting $\varepsilon$ to higher values). On the other hand, upon interaction with SUT instances that are similar to the previous ones, the parameter $\varepsilon$ is adjusted to increase exploitation (setting $\varepsilon$ to lower values). SaFReL detects the similarity between SUT instances by calculating \textit{cosine similarity} between the performance sensitivity vectors of SUT instances, as shown in Eq. \ref{eq:Similarity}.  
\begin{equation} \label{eq:Similarity}
\begin{split}
\textrm{similarity}(k,k-1)&=\frac{SV^k\ SV^{k-1}}{\|SV^k\|\|SV^{k-1}\|}\\
&=\frac{\sum_{i=1}^{3} {SV_i^{k}SV_i^{k-1}}}{\sqrt{\sum_{i=1}^{3}{(SV_i^{k})}^2}\sqrt{\sum_{i=1}^{3}{(SV_i^{k-1})}^2}}
\end{split}
\end{equation}
where \(SV^k\) represents the sensitivity vector of the \(k^{th}\) SUT instance and \(SV_i^k\) represents the \(i^{th}\) element of vector \(SV^k\). \blue{The sensitivity vector contains the values of the sensitivity indicators of the SUT instance, \(Sen^C\), \(Sen^M\) and \(Sen^D\). The  performance sensitivity indicators assume values in the range $[0, 1]$ and represent the sensitivity degree of the SUT to CPU, memory and disk respectively. Their values could be set empirically or even intuitively, and SaFReL uses the approximate estimated similarity to tune the $\varepsilon$ value adaptively (See Section \ref{sec:: Experiments_and_results}).}\\ 

\textit{\textbf{Reward Signal.}} The agent receives a reward signal indicating the effectiveness of the applied action in each learning step to guide the agent toward reaching the intended performance breaking point. We derive a utility function as a weighted linear combination of two functions indicating the response time deviation and resource usage, which is as follows:
\blue{
\begin{equation} \label{eq:reward}
R_n=\beta U_n^r+(1-\beta)U_n^E						\end{equation}}
where \(U_n^r\) represents the deviation of response time from the response time requirement, \(U_n^E\) indicates the resource usage, and \(\beta\), \(0\leq\beta\leq1\) is a parameter intended to prioritize different aspects of stress conditions, i.e., response time deviation or limited resource availability. \(U_n^r\) is defined as follows:
\begin{equation} \label{eq:Un}
U_n^r = 
     \begin{cases}
       \text{0,} &\quad RT_n^\prime\leq RT^q\\
       \frac{(RT_n^\prime-RT^q)}{(RT^b-RT^q)}, &\quad RT_n^\prime > RT^q\\
     \end{cases}							
\end{equation}

where $RT_n^\prime$ is the measured response time, \(RT^q\) is the response time requirement and \(RT^b\) is the threshold defining the performance breaking point. \(U_n^E\) represents the resource utilization in the reward signal, and is a weighted combination of the resource utilization values. It is defined using the following equation:
\begin{equation} \label{eq:UE}
U_n^E= Sen^C CUI_n^{\prime}+ Sen^M MUI_n^{\prime} +Sen^D DUI_n^{\prime}					
\end{equation}
where \(CUI_n^\prime\), \(MUI_n^\prime\), and \(DUI_n^\prime\) represent CPU, memory and disk utilization improvements respectively, and \(Sen^C\), \(Sen^M\) and \(Sen^D\) are the performance sensitivity indicators of the SUT, and assume values in the range $[0, 1]$.

\section{Performance Testing using Self-Adaptive Fuzzy Reinforcement Learning} \label{sec::Stress Testing using Self-Adaptive}
In this section, we describe details of the procedure of SaFReL to generate the performance test cases resulting in reaching the performance breaking points for various types of SUTs. The tester agent learns how to generate the target test cases for different types of software without access to source code or system models. The procedure of SaFReL, which includes initial and transfer learning phases, is as follows:

The agent measures the quality parameters and identifies the state- membership degree pair \((S_n,\mu_n )\) through the fuzzy state detection, where \(S_n\) is the fuzzy state of the environment and \(\mu_n\) indicates the membership degree, which means to what extent the environment has assumed that state. Then, according to the action selection strategy, the agent selects one action, \(a_n \in A_n\) based on the previously learned policy or through exploring the state action space. The agent takes the selected action and executes the SUT. In the next step the agent detects the new state of the SUT, \((S_{n+1},\mu_{n+1})\) and receives a reward signal, \(r_{n+1}\in \mathbb{R}\), indicating effectiveness of the applied action. After detecting the new state and receiving the reward, it updates the stored experience (learned policy). The whole procedure is repeated until meeting the stopping criterion that is reaching the performance breaking point, \((RT^b)\).
The experience of the agent is defined in terms of the policy that the agent learns. A policy is a mapping between each state and action and specifies the probability of taking action \(a\) in a given state \(s\). The purpose of the agent in the learning is to find a policy that maximizes the expected long-term reward achieved over the further learning trials, which is formulated as follows: \cite{sutton2018reinforcement}:
\begin{equation} \label{eq:purposeofRL}
R_n=r_{n+1}+\gamma r_{n+2}+...+\gamma^k r_{n+k+1}= \sum_{k=0}^{\infty} \gamma^k r_{n+k+1}
\end{equation}

where \(\gamma\) is a discount factor specifying to what extent the agent prioritize future rewards compared to the immediate one. We use Q-learning as a model-free RL algorithm in our framework. In Q-Learning, a utility value \(Q^\pi (s,a)\) is assigned to each pair of state and action, which is defined as follows: \cite{sutton2018reinforcement}:
\begin{equation} \label{eq:Q value}
Q^\pi (s,a)=E^\pi [R_n | s_n=s,a_n=a]
\end{equation}

The q-values, \(Q^\pi (s,a)\), form the experience base of the agent, on which the agent relies for the action selection. The q-values are updated incrementally during the learning. According to using fuzzy state modeling, we include the membership degree of the detected state of the environment, $\mu_n^s$, in the typical updating equation of q-values to take into account the impact of the uncertainty associated with the fuzzy state, which is as follows:
\begin{equation} \label{eq:Q updating}
Q(s_n,a_n)=\mu_n^s [(1-\alpha) Q(s_n,a_n)+ \alpha(r_{n+1}+ \gamma \max\limits_{a^{\prime}} Q(s_{n+1},a^{\prime}))]
\end{equation}
where \(\alpha\), \(0 \leq \alpha \leq 1\) is the learning rate, which adjusts to what extent the new utility values affect (overwrite) the previous q-values. Finally, the agent finds the optimal policy to reach the target, which suggests the action maximizing the utility value for a given state \(s\): 
\begin{equation} \label{eq:action_from_learnt_policy}
a(s)= \argmax\limits_{a^{\prime}} Q(s,a^{\prime})
\end{equation}								
The agent selects the action based on Eq. \ref{eq:action_from_learnt_policy} when it is supposed to exploit the learned policy.
SaFReL implements two learning phases: initial and transfer learning.

\textbf{\textit{Initial learning.}} Initial learning occurs during the interaction with the first SUT instance. The initial convergence of the policy takes place upon the initial learning. The agent stores the learned policy (in terms of a table containing q-values, Q-table). It repeats the learning episode multiple times on the first SUT instance to achieve the initial convergence of the policy.

\textbf{\textit{Transfer learning.}} SaFReL goes through the transfer learning phase, after the initial convergence. During this phase, the agent uses the learned policy upon observing SUT instances with similar performance sensitivity to the previously observed ones, while keeping the learning running, i.e., updating the policy upon detecting new SUT instances with different performance sensitivity. Strategy adaptation is used in the transfer learning phase and makes the agent adapt to various SUT instances. Algorithms \ref{Algorithm: SaFReL} and \ref{Algorithm: Fuzzy Q-Learning} present the procedure of SaFReL in both initial learning and transfer learning phases.

\begin{algorithm}[H]
\SetAlgoLined
\caption{SaFReL: Self-adaptive Fuzzy Reinforcement Learning-based Performance Testing}\label{Algorithm: SaFReL}
\textbf{Required:} $S, A, \alpha, \gamma$;\\
\textrm{Initialize q-values,\  \(Q(s,a)= 0\ \forall s \in \mathbb{S},\  \forall a \in \mathbb{A}\ \textrm{and}\ \epsilon=\upsilon\ ,0 < \upsilon <1\)};\\
\textrm{Observe the first SUT instance};\\ 
\Repeat{initial convergence}{
        Fuzzy Q-Learning (with initial action selection strategy, e.g. $\epsilon$-greedy, initialized $\epsilon$)\;
    }
\textrm{Store the learnt policy};\\
\textrm{Start the transfer learning phase};\\
\While{true}{
Observe a new SUT instance;\\
Measure the similarity;\\
Apply strategy adaptation to adjust the degree of exploration and exploitation (e.g. tuning parameter $\epsilon$ in $\epsilon$-greedy);\\
Fuzzy Q-Learning with adapted strategy (e.g. new value of $\epsilon$);\\
}
\end{algorithm}

\begin{algorithm}[H]
\SetAlgoLined
\caption{Fuzzy Q-Learning}\label{Algorithm: Fuzzy Q-Learning}
\Repeat{meeting the stopping criteria (reaching performance breaking point) }{
        \textrm{1. Detect the fuzzy state-degree pair \((S_n,\mu_n)\) of the SUT};\\ 
        \textrm{2. Select an action using the action selection strategy (e.g. $\epsilon$-greedy: select $a_n= \argmax_{a\in \mathbb{A}} Q(s_n,a)$ with probability (1-$\epsilon$) or a random $a_k$, $a_k \in \mathbb{A}$ with probability $\epsilon$)};\\
        3. Take the selected action, execute the SUT;\\
        4. Detect the new fuzzy state-degree $(S_{n+1},\mu_{n+1})$ of the environment;\\
        5. Receive the reward signal, $R_{n+1}$;\\
        6. Update the q-value of the pair of previous state and applied action\\ 
        $Q(s_n,a_n)=\mu_n^s [(1-\alpha) Q(s_n,a_n)+ \alpha(r_{n+1}+ \gamma \max\limits_{a^{\prime}} Q(s_{n+1},a^{\prime}))]$\;	
    }
\end{algorithm}

\section{Evaluation} \label{sec::Evaluation}
In this section, we present the experimental evaluation of the proposed self-adaptive fuzzy RL-based performance testing framework, SaFReL. We assess the performance of SaFReL, in terms of efficiency in generating the performance test cases and adaptivity to various types of SUT programs, i.e., how well it can adapt its functionality to new cases while preserving its efficiency. \blue{ Therefore, we examine the efficiency of SaFReL (in the transfer learning phase) compared to a typical testing process for this target, which involves generating the performance test cases through changing the availability of the resources based on the defined actions in an exploratory (random) way, which is called \textit{typical stress testing} hereafter.} 
We also evaluate the sensitivity of SaFReL to the learning parameters. The goal of the experimental evaluation is to answer the following research questions:
\begin{itemize}
\item RQ1. How efficiently can SaFReL generate the test cases leading to the performance breaking points for different software programs compared to a typical testing procedure?
\item RQ2. How adaptively can SaFReL act on various software programs with different performance sensitivity?
\item RQ3. How is the efficiency of SaFReL affected by changing the learning parameters?
\end{itemize}
The following sub-sections describe the proposed setup for conducting the experiments, the evaluation metrics, and the analysis scenarios designed for answering the above research questions. 

\subsection{Experiments Setup} \label{sec:: Experiments Setup}
\begin{figure}[ht]
\centering
\includegraphics[width=.98\textwidth]{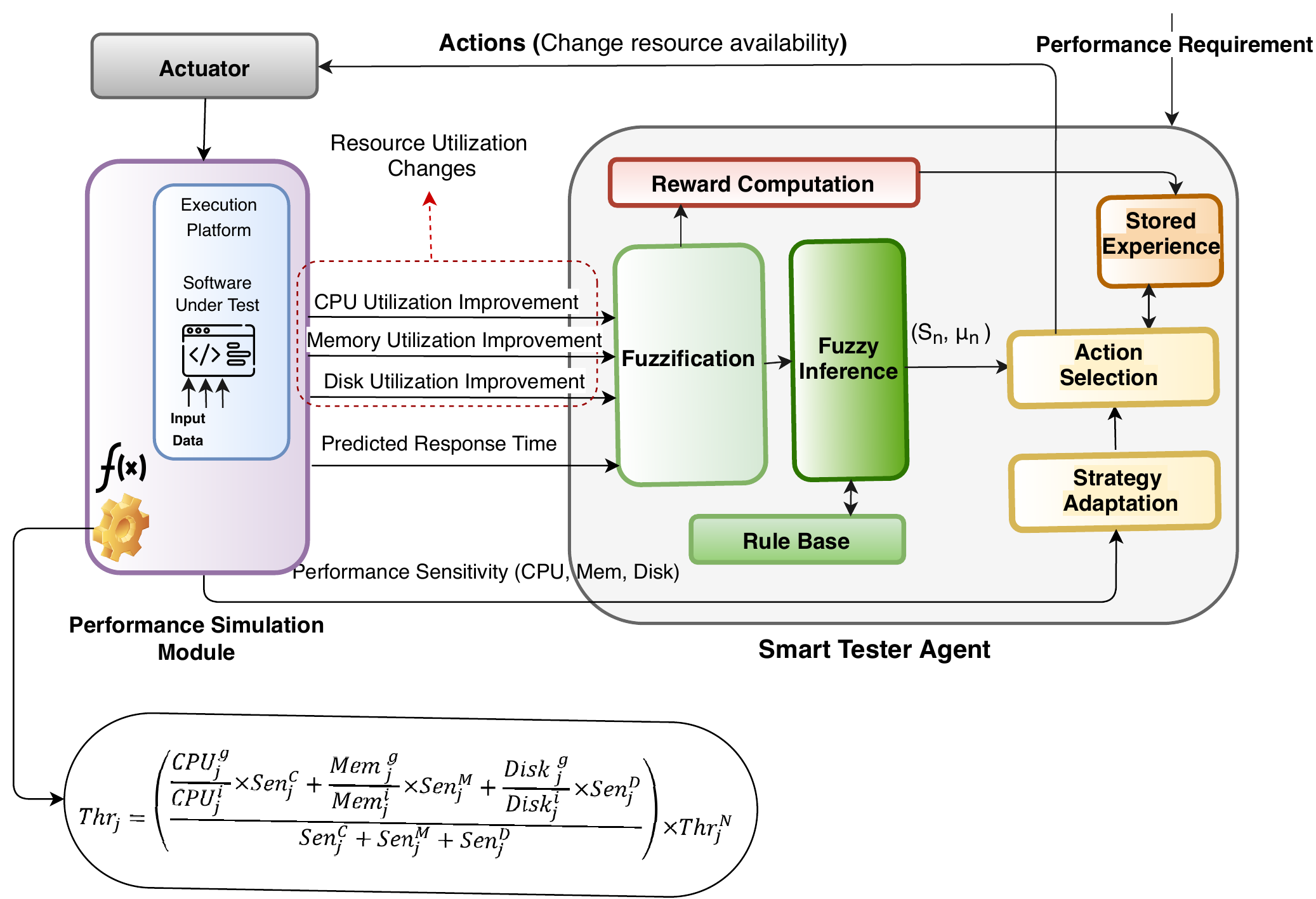}
\caption{\blue{Implementation structure} }
\label{fig: Our implementation structure}
\end{figure}

In this study, we implement the proposed smart testing framework (agent) along with \blue{ a performance simulation module simulating the performance behavior of SUT programs under different execution conditions.} The simulation module receives the resource sensitivity values and based on the amounts of resources demanded initially and the amounts of them granted after taking each action, estimates the program throughput using the following equation proposed by \cite{taheri2016vmbbthrpred}:
\begin{equation} \label{eu_eqn}
Thr_j=\frac{\frac{CPU_j^g}{CPU_j^i}Sen_j^C +\frac{Mem_j^g}{Mem_j^i}Sen_j^M+ \frac{Disk_j^g}{Disk_j^i}Sen_j^D} {Sen_j^C+ Sen_j^M+ Sen_j^D}\times Thr_j^N
\end{equation}
where \(CPU_j^i\), \(Mem_j^i\) and \(Disk_j^i\) the amounts of CPU, memory and disk resources demanded by program j at the initial state and \(CPU_j^g\), \(Mem_j^g\) and \(Disk_j^g\) are the amounts of resources granted to program j after taking an action, which modifies the resource availability. \(Sen_j^C\), \(Sen_j^M\) and \(Sen_j^D\) represent the CPU, memory and disk sensitivity values of program j, and \(Thr_j^N\) represents the nominal throughput of program j in an isolated, contention free environment. The response time of the program is calculated as \(RT_j=\frac{1}{Thr_j}\) in the simulation module.
Figure \ref{fig: Our implementation structure} presents the implementation structure including SaFReL along with the implemented performance simulation module.\blue{ In our implementation, the performance simulation module simulates the performance behavior of the SUT program and the testing agent interacts with the simulation module to capture the quality measures used for state detection.}

Table \ref{table:Programs and sen values} shows the list of programs and the corresponding resource sensitivity values used in the experimentation, the table data obtained from \cite{taheri2016vmbbthrpred}. The collection listed in Table \ref{table:Programs and sen values} includes various CPU-intensive, memory-intensive and disk-intensive types of programs and also the programs with combined types of resource sensitivity.
The SUTs are instances of the programs listed in Table \ref{table:Programs and sen values} and are characterized with various initial amounts of resources and also different values of response time requirements. 
Two analysis scenarios are designed to answer the evaluation research questions. The first one focuses on efficiency and adaptivity evaluation of the framework on various SUTs. In the second analysis scenario, the sensitivity of the approach to changes of the learning parameters are studied. The efficiency and adaptivity are measured (evaluated) according to following specification: 
\begin{itemize}
\item \textit{Efficiency} is measured in terms of number of learning trials required by the tester agent to achieve the testing target, which is reaching the intended performance breaking point. Number of learning trials is an indicator of the required computation time to generate the proper test case leading to the performance breaking point.  

\item \textit{Adaptivity} is evaluated in terms of number of additional learning trials (computation time) required to re-adapt the learned policy to new observations for achieving the target.
 
\end{itemize}

\begin{table}[h!]
\centering
\caption{Programs and the corresponding sensitivity values used for experimental evaluation \cite{taheri2016vmbbthrpred}}
\begin{tabu} to \textwidth{ | X[l] | X[l] | } 
 \hline
 \textbf{Programs} & \textbf{Resource Sensitivity Values (\(Sen^C\), \(Sen^M\) and \(Sen^D\))} \\ 
 \hline
 Build-apache & (0.96, 0.04, 0.00) \\
 \hline
 n-queens & (0.97, 0.00, 0.00) \\
 \hline
 John-the-ripper &(0.96, 0.00, 0.00)\\
 \hline
 Apache & (0.97, 0.03, 0.00)\\
 \hline
 Dcraw & (0.48, 0.04, 0.00)\\
 \hline
 X264 & (0.41, 0.02, 0.00)\\
 \hline
 Unpack-linux & (0.18, 0.09, 0.35)\\
 \hline
 Build-php & (0.97, 0.07, 0.00)\\
 \hline
 Blogbench & (0.11, 0.81, 0.18)\\
 \hline
 Bork & (0.00, 0.53, 0.20)\\
 \hline
 Compress-gzip & (0.00, 0.00, 0.47)\\
 \hline
 Aio-stress & (0.00, 0.30, 0.80)\\
 \hline
\end{tabu}
\label{table:Programs and sen values}
\end{table}

\subsection{Experiments and Results} \label{sec:: Experiments_and_results}
\subsubsection{Efficiency and Adaptivity Analysis}
To answer RQ1 and RQ2, the performance of SaFReL is evaluated based on its efficiency in generating the performance test cases leading to the performance breaking points of different SUTs and its adaptation capability to new SUTs with performance sensitivity different from previously observed ones. 
We select two sets of SUT instances: i) one including SUTs similar in the aspect of performance sensitivity to resources, i.e., similar with regard to the primarily demanded resource (homogenous SUTs); and ii) the other set contains SUT instances different in performance sensitivity (heterogeneous SUTs). The SUT instances assume different initial amounts of CPU, memory and disk resources, and response time requirements. The amounts of resources, CPU, memory and disk capacity, were initialized with different values in the range [1, 10] cores, [1, 50] GB, [100, 1000] GB respectively. The response time requirements range from 500 to 3000 ms. The intended performance breaking point for the SUT instances is defined as the point in which the response time exceeds 1.5 times the response time requirement.

In the efficiency analysis, we set the learning parameters, learning rate and discount factor, to $0.1$ and $0.5$, respectively. We study the impacts of different variants of $\varepsilon$-greedy algorithm as the action selection strategy on the efficiency and adaptivity of the approach during the analysis. We investigate three variants of $\varepsilon$-greedy with $\varepsilon=0.2$, $\varepsilon=0.5$, and decaying $\varepsilon$, and also the proposed adaptive $\varepsilon$ selection method.      

\begin{figure}[h]
\centering
\includegraphics[width=.98\textwidth, height=9cm]{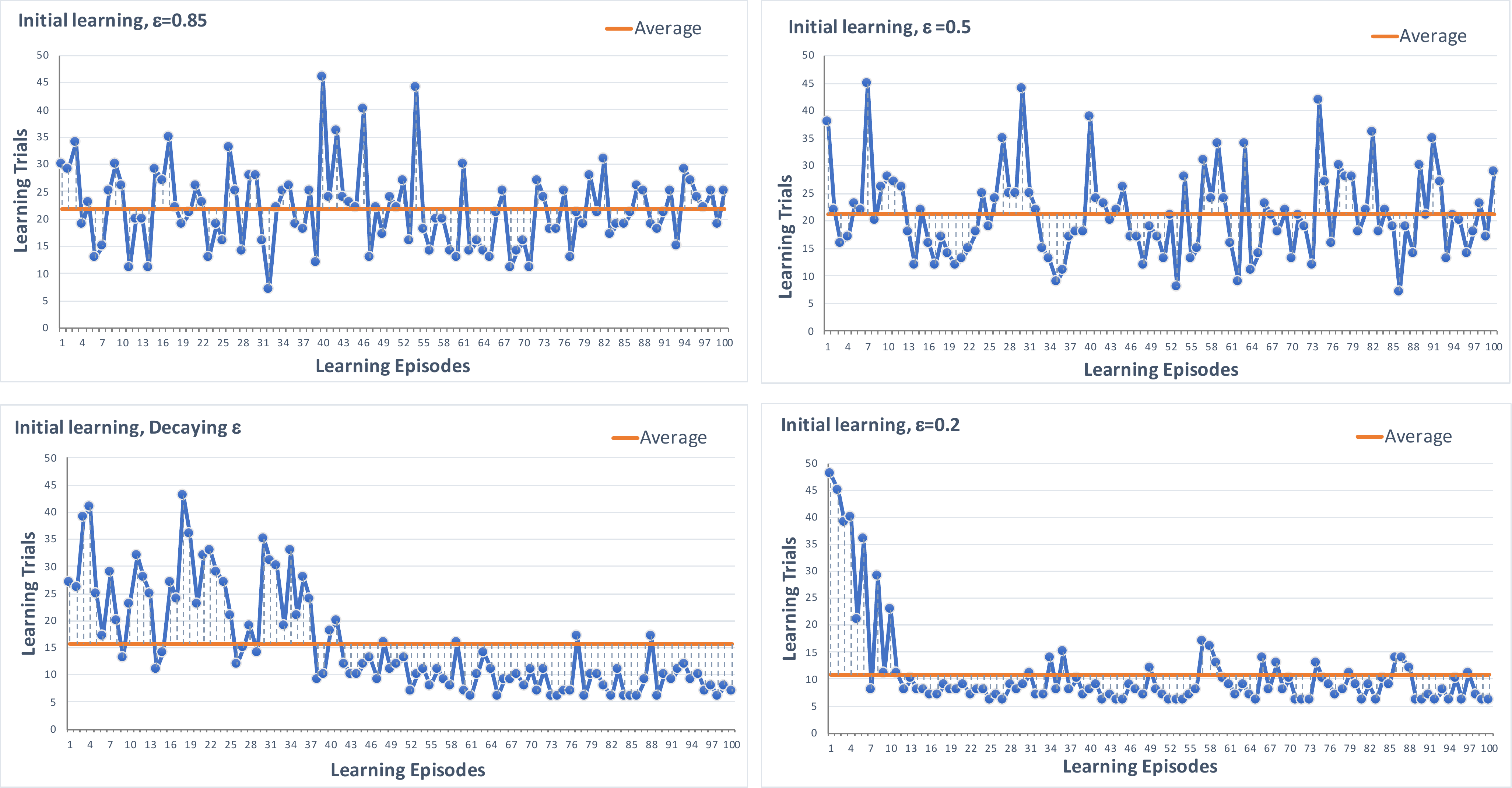}
\caption{Initial convergence of SaFReL in 100 learning episodes during the initial learning }
\label{fig: Efficiency of SaFReL- initial learning}
\end{figure}

\textit{Learning setup.} First, we need to set up the initial learning. For choosing a proper configuration for the action selection strategy in the initial learning, we evaluate the performance of different variants of $\varepsilon$-greedy algorithm, in terms of the number of required learning trials for initial convergence (Figure \ref{fig: Efficiency of SaFReL- initial learning}). For the initial convergence, we run the initial learning on the first SUT 100 times, namely 100 learning episodes. \blue{ Table \ref{table: Efficiency of SaFReL- initial learning} presents a quick summarized view of the average learning trials during the last 10 episodes that are considered as the achieved values upon the convergence of the initial learning.}
\blue{As shown in Figure \ref{fig: Efficiency of SaFReL- initial learning} and Table \ref{table: Efficiency of SaFReL- initial learning}, using $\varepsilon$-greedy with $\varepsilon=0.2$ results in the fastest initial convergence, which has also led to the lowest number of trials compared to the other variants of $\varepsilon$-greedy. The number of learning trials after about 10 episodes starts converging and during the last 10 episodes it converges to approximately 7 trials.}

\begin{table}[b]
\begin{center}
\caption{\blue{Initial convergence of SaFReL in the initial learning regarding using different variants of action selection strategy}}
\begin{tabular}{ | m{4.5cm} | m{1.3cm}| m{1.3cm}|m{1.3cm}| m{1.35cm} | } 
\hline
\textbf{}& \multicolumn{4}{|c|}{SaFReL - Initial Learning} \\
\hline
\textbf{\small Action Selection Strategy: $\epsilon$-greedy} & $\epsilon=0.85$ & $\epsilon=0.5$ & $\epsilon=0.2$ & \small decaying $\epsilon$  \\ 
\hline
\textbf{\blue{\small Number of learning trials (after convergence)}} & $22$ & $21$ & $7$ & $9$ \\ 
\hline
\end{tabular}
\label{table: Efficiency of SaFReL- initial learning}
\end{center}
\end{table}
Once the initial convergence occurs, SaFReL is ready to act on various SUTs and is expected to be able to reuse the learned policy to meet the intended performance breaking points on further SUT instances, while still keeping the learning running. 
The optimal policy learned in the initial learning is not influenced by the used action selection strategy, since Q-learning is an off-policy learning algorithm \cite{sutton2018reinforcement}. It implies that the learner finds the optimal policy independently of how the actions have been selected (action selection strategy). For the sake of efficiency, we choose the one that resulted in the fastest convergence.

In the following sections, first, we investigate the efficiency of SaFReL compared to a typical stress testing procedure, when acting on homogeneous and heterogeneous sets of SUTs, then its capability to adapt to new SUTs with different performance sensitivity. 

\begin{figure}[h]
\centering
\includegraphics[width=.98\textwidth, height=9cm]{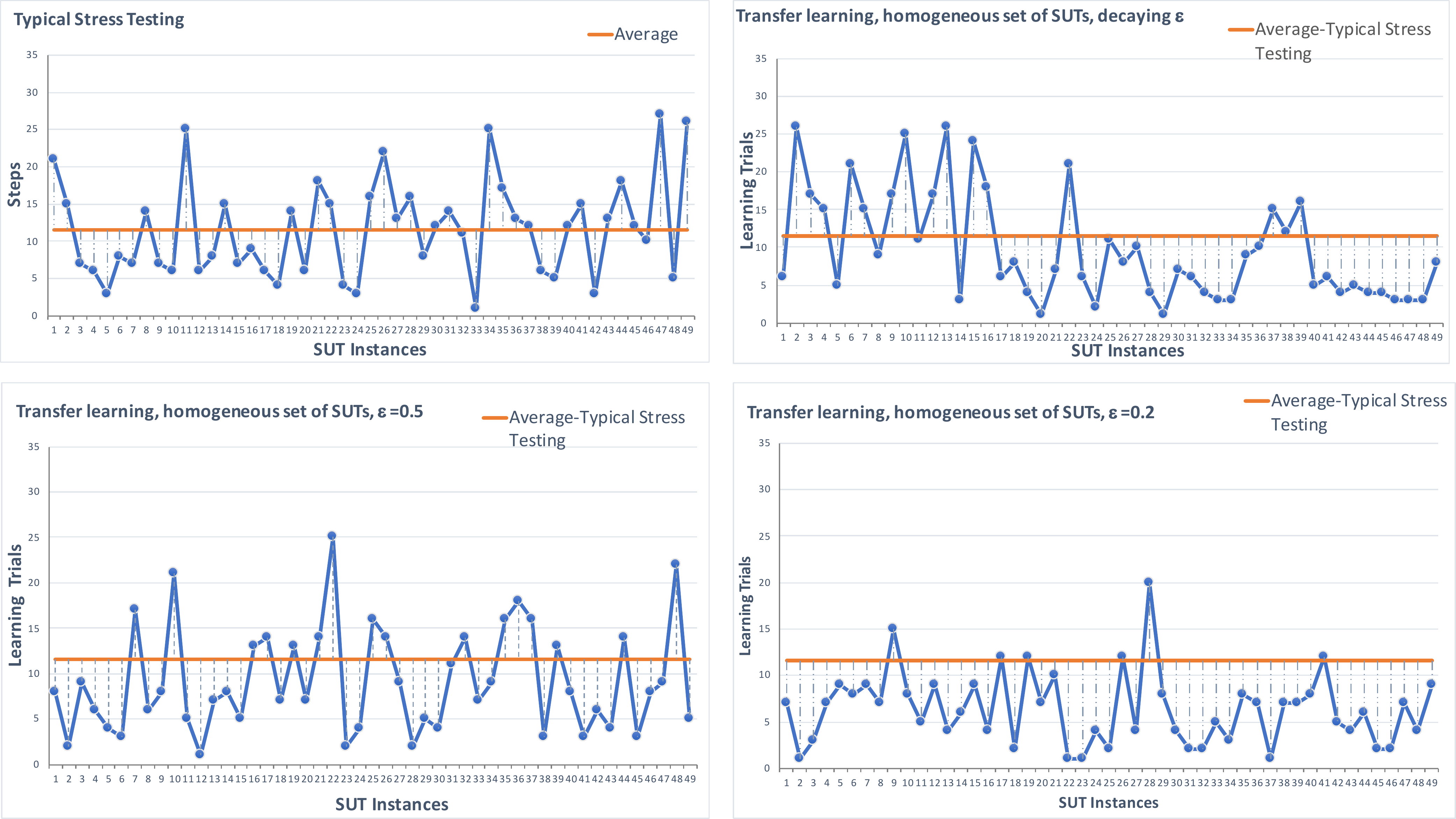}
\caption{Efficiency of SaFReL on a homogeneous set of SUTs in the transfer learning }
\label{fig: Efficiency of SaFReL-homogeneous set of SUTs-transfer learning}
\end{figure}

\paragraph{\textit{I. Homogeneous set of SUTs.}} We select CPU-intensive programs and make a homogeneous set of SUT instances during our analysis in this step. We simulate the performance behavior of 50 instances of the CPU-intensive programs, Build-apache, n-queens, John-the-ripper, Apache, Dcraw, Build-php, X264, and vary both the initial amounts of resources granted and the response time requirements. Figure \ref{fig: Efficiency of SaFReL-homogeneous set of SUTs-transfer learning} shows the efficiency of SaFReL on a homogeneous set of CPU-intensive SUTs compared to a typical stress testing procedure regarding using $\varepsilon$-greedy with different values of $\varepsilon$. Table \ref{table: Efficiency of SaFReL-homogeneous set of SUTs-transfer learning} presents the average number of trials/steps for generating the target performance test case in the proposed approach and the typical testing procedure. As shown in Figure \ref{fig: Efficiency of SaFReL-homogeneous set of SUTs-transfer learning}, it keeps the number of required trials for $\approx 94\%$ of the SUTs below the average number of required steps in the typical stress testing. 
Table \ref{table: improvement-SaFReL-homogeneous set of SUTs-transfer learning} shows the resulting improvement in the average number of required trials/steps for meeting the target, which implies reduction in the required computation time, compared to the typical stress testing process.

In the transfer learning, the agent reuses the learned policy based on the allowed degree of policy reusing according to its action selection strategy in the transfer learning. As shown in Table \ref{table: Efficiency of SaFReL-homogeneous set of SUTs-transfer learning}, 
it implies that in the transfer learning the agent does fewer trials (based on the degree of allowed policy reusing) to meet the target on new cases, which leads to a higher efficiency. According to Table \ref{table: improvement-SaFReL-homogeneous set of SUTs-transfer learning}, on a homogeneous set of SUTs, more policy reusing leads to higher efficiency (more computation time improvement). 

\begin{table}[h]
\begin{center}
\caption{Average number of trials/steps for generating the target performance test case on the homogeneous set of SUTs}
\begin{tabular}{ | m{3.2cm} | m{1.5cm}| m{1.5cm}|m{1.5cm}| m{2cm} |  } 
\hline
\textbf{}& \multicolumn{3}{|c|}{SaFReL with $\epsilon$-greedy}& \\
\hline
\textbf{\small Approach} & $\epsilon=0.5$ & \small decaying $\epsilon$ & $\epsilon=0.2$ & Typical stress testing  \\ 
\hline
\textbf{\small Average number of trials/steps} & $10$ & $10$ & $7$ & $12$ \\ 
\hline
\end{tabular}
\label{table: Efficiency of SaFReL-homogeneous set of SUTs-transfer learning}
\end{center}
\end{table} 

\begin{table}[h]
\begin{center}
\caption{Computation time improvement on the homogeneous set of SUTs}
\begin{tabular}{ | m{5.9cm} | m{1.45cm}| m{1.45cm}| m{1.45cm} |  } 
\hline
\textbf{}& \multicolumn{3}{|c|}{SaFReL} \\
\hline
\textbf{\small Action Selection Strategy: $\epsilon$-greedy} & $\epsilon=0.5$ & \small decaying $\epsilon$ & $\epsilon=0.2$   \\ 
\hline
\textbf{\small Improvement in the number of trials} & $16\%$ & $16\%$ & $42\%$  \\ 
\hline
\end{tabular}
\label{table: improvement-SaFReL-homogeneous set of SUTs-transfer learning}
\end{center}
\end{table} 

\begin{figure}[h]
\centering
\includegraphics[width=.98\textwidth, height=9cm]{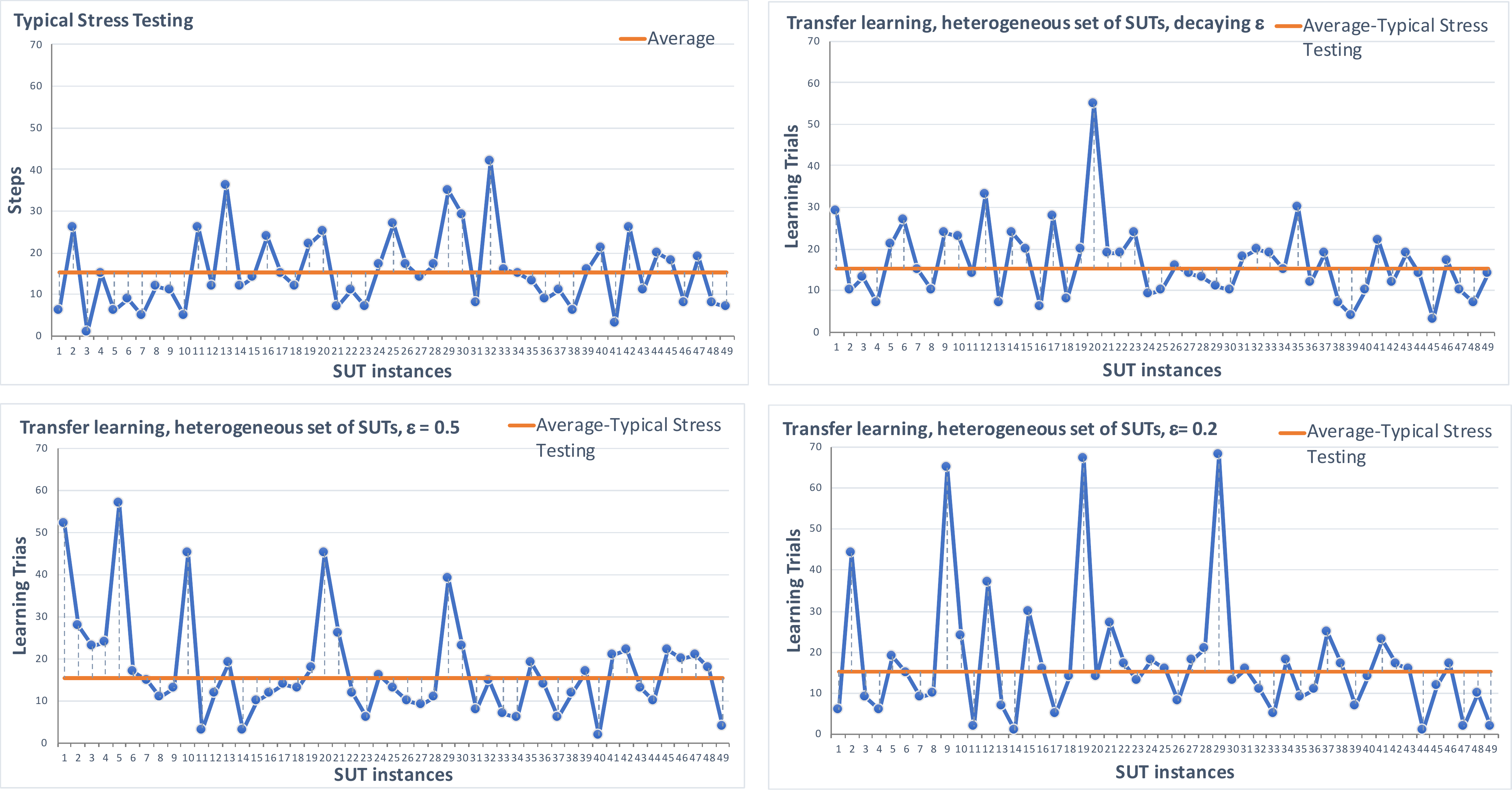}
\caption{Efficiency of SaFReL on a heterogeneous set of SUTs regarding the use of typical configurations of $\epsilon$-greedy}
\label{fig:Efficiency of SaFReL-heterogeneous set of SUTs-typical epsilon}
\end{figure}

\paragraph{\textit{II. Heterogeneous set of SUTs.}} In this part of the analysis, to complete the answer to RQ1 and and also answer RQ2, we examine the efficiency and adaptivity of SaFReL during the transfer learning on a heterogeneous set of SUTs including various CPU-intensive, memory-intensive and disk-intensive ones. We simulate the performance behavior of 50 SUT instances from the list of the programs in Table \ref{table:Programs and sen values}. We evaluate the efficiency of SaFReL on the heterogeneous set of SUTs compared to the typical stress testing procedure regarding using $\varepsilon$-greedy with $\varepsilon=0.2$, $0.5$, and decaying $\varepsilon$ (Figure \ref{fig:Efficiency of SaFReL-heterogeneous set of SUTs-typical epsilon}). As shown in Figure \ref{fig:Efficiency of SaFReL-heterogeneous set of SUTs-typical epsilon} the transfer learning algorithm with a typical configuration of the action selection strategy, such as $\varepsilon=0.2$, $0.5$ and decaying $\varepsilon$, which imposes a certain degree of policy reusing based on the value of $\varepsilon$ does not work well. It does not outperform the typical stress testing, but also slightly degrades in some cases of $\varepsilon$. When the smart agent acts on a heterogeneous set of SUTs, blind replaying of the learned policy (i.e., just based on the value of $\varepsilon$) is not effective, and the tester agent needs to know where it should do policy reusing and where it requires more exploration to update the policy. 

As described in Section \ref{sec::Adaptive Action Selection and Reward}, to solve this issue and improve the performance of SaFReL when it acts on a heterogeneous set of SUTs, it is augmented with a simple meta-learning feature enabling it to detect the heterogeneity of the SUT instances and adjust the value of parameter $\varepsilon$, adaptively. In general, it implies that when the smart tester agent observes a SUT instance different from the previously observed ones wrt the performance sensitivity, it changes the action selection strategy to doing more exploration and upon detecting a SUT instance with the same performance sensitivity as the previous ones, it makes the action selection strategy strive for more exploitation. As illustrated in Section \ref{sec::Adaptive Action Selection and Reward}, the strategy adaptation module, which fulfills this function, measures the similarity between SUTs at two levels of observations, then based on the measured values, adjusts the value of parameter $\varepsilon$. The threshold values of similarity measures and the adjustments for parameter $\varepsilon$ in the experimental analysis are described in Algorithm \ref{Algorithm:Adaptive selection}.

\begin{algorithm}[H]
\caption{Adaptive $\epsilon$ selection}\label{Algorithm:Adaptive selection}
\begin{algorithmic}
    \IF{$similarity_{k,k-1}\geq 0.8$} 
        \IF{$similarity_{k,k-2}\geq 0.8$}
            \STATE $\epsilon\gets 0.2$
        \ELSE
            \STATE $\epsilon\gets 0.5$
        \ENDIF
    \ELSIF{$similarity_{k,k-1}< 0.8$}
        \STATE $\epsilon\gets 0.5$
    \ENDIF 
\end{algorithmic}
\end{algorithm}

\begin{figure}[h]
\centering
\includegraphics[width=.78\textwidth, height=5cm]{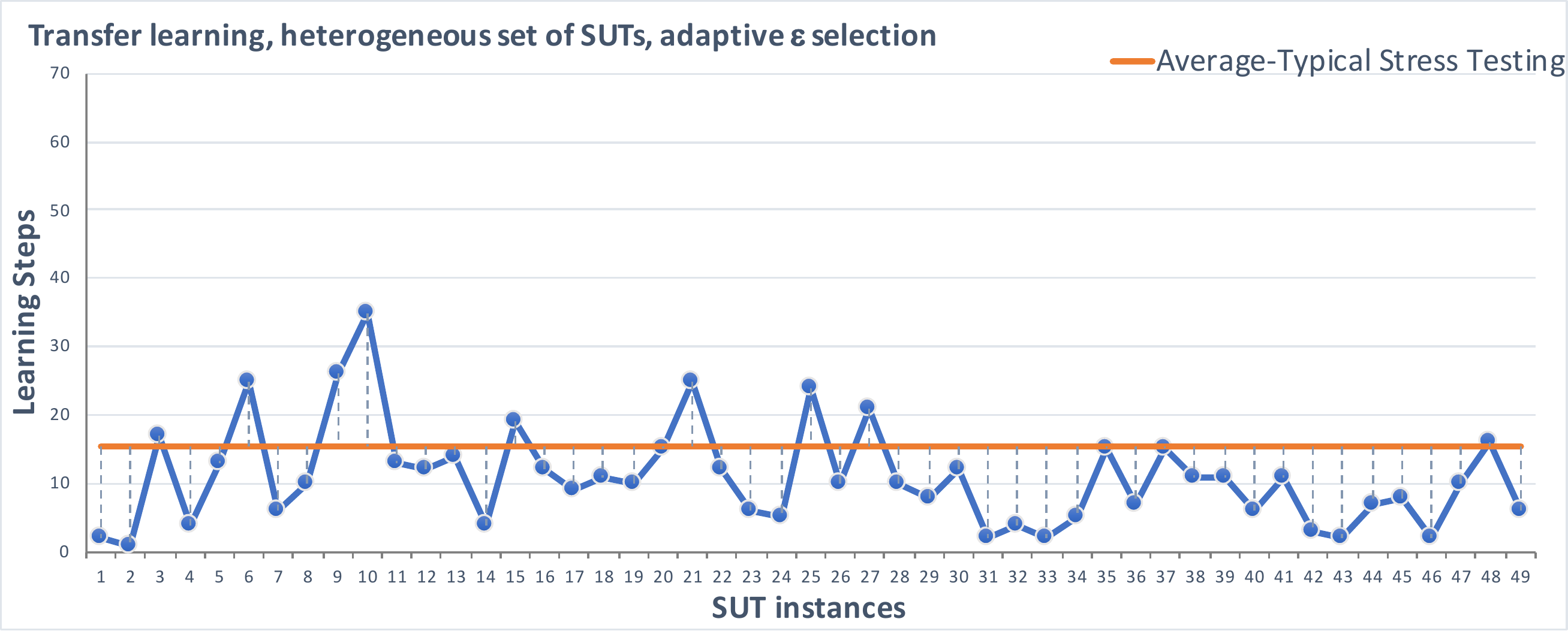}
\caption{Efficiency of SaFReL on a heterogeneous set of SUTs regarding the use of adaptive $\epsilon$-greedy action selection strategy }
\label{fig:Efficiency of SaFReL-heterogeneous set of SUTs-adaptive}
\end{figure}

Figure \ref{fig:Efficiency of SaFReL-heterogeneous set of SUTs-adaptive} shows the efficiency of SaFReL regarding the use of similarity detection and the adaptive $\varepsilon$-greedy action selection strategy on a heterogeneous set of SUTs. Regarding the use of adaptive $\varepsilon$ selection, SaFReL makes a considerable improvement and is able to keep the number of required trials for reaching the target on approximately $\approx 82\%$ of SUTs below the corresponding average value in the typical stress testing. Meanwhile, the average number of learning trials is totally lower than the typical stress testing procedure.
Table \ref{table: Efficiency of SaFReL-heterogeneous set of SUTs-transfer learning} presents the average number of trials/steps for generating the target performance test case in SaFReL and the typical stress testing when they act
on a heterogeneous set of SUTs. Table \ref{table: improvement-SaFReL-heterogeneous set of SUTs-transfer learning} shows the corresponding resulting improvement in the computation time respectively.

\begin{table}[h]
\begin{center}
\caption{Average number of trials/steps for generating the target performance test case on the heterogeneous set of SUTs}
\begin{tabular}{ | m{2.3cm} | m{1.4cm}| m{1.4cm}|m{1.4cm}|m{1.4cm}| m{1.5cm} |  } 
\hline
\textbf{}& \multicolumn{4}{|c|}{SaFReL with $\epsilon$-greedy}& \\
\hline
\textbf{\small Approach} & $\epsilon=0.5$ & \small decaying $\epsilon$ & $\epsilon=0.2$ & adaptive $\epsilon$ & Typical stress testing  \\ 
\hline
\textbf{\small Average number of trials/steps} & $18$ & $17$ & $18$ & $11$ & $16$ \\ 
\hline
\end{tabular}
\label{table: Efficiency of SaFReL-heterogeneous set of SUTs-transfer learning}
\end{center}
\end{table} 
\begin{table}[h]
\begin{center}
\caption{Computation time improvement on the heterogeneous set of SUTs}
\begin{tabular}{ | m{4cm} | m{1.45cm}| m{1.45cm}| m{1.45cm}|m{1.45cm}| } 
\hline
\textbf{}& \multicolumn{4}{|c|}{SaFReL} \\
\hline
\textbf{\small Action Selection Strategy: $\epsilon$-greedy} & $\epsilon=0.5$ & \small decaying $\epsilon$ & $\epsilon=0.2$ & adaptive $\epsilon$ \\ 
\hline
\textbf{\small Improvement in the number of trials} & No & No & No & $31\%$ \\ 
\hline
\end{tabular}
\label{table: improvement-SaFReL-heterogeneous set of SUTs-transfer learning}
\end{center}
\end{table} 

\begin{figure}[h]
\centering
\includegraphics[width=.98\textwidth, height=9cm]{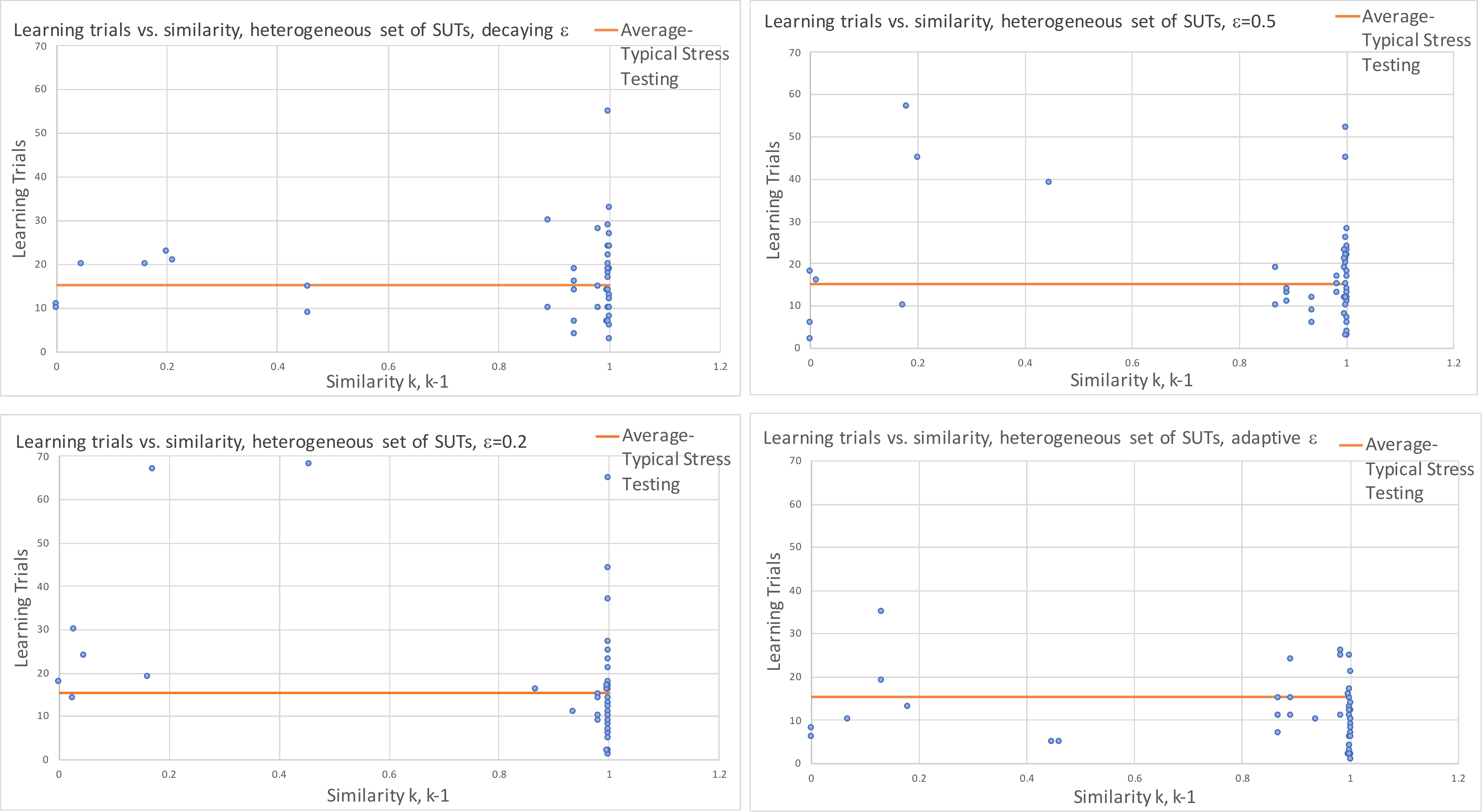}
\caption{Adaptivity of SaFReL on a heterogeneous set of SUTs regarding the use of different variants of action selection strategy}
\label{fig:Adaptivity of SaFReL-heterogeneous set of SUTs}
\end{figure}

To answer RQ2, we investigate the adaptivity of SaFReL on the heterogeneous set of SUTs regarding the use of different variants of action selection strategy including adaptive $\varepsilon$ selection (Figure \ref{fig:Adaptivity of SaFReL-heterogeneous set of SUTs}). As shown in Figure \ref{fig:Adaptivity of SaFReL-heterogeneous set of SUTs}, the number of  required learning trials versus detected similarity is used to depict how adaptive SaFReL can act on a heterogeneous set of SUTs regarding the use of different configurations of $\varepsilon$. It shows that SaFReL with adaptive $\varepsilon$ is able to adapt to changing situations, e.g., a mixed heterogeneous set of SUTs. In other words, on around $\approx 75\%$ of SUTs that are completely different from the previous ones (i.e., with $similarity_{k,k-1} < 0.8$) it still keeps the number of required trials to meet the target below the average value of the typical stress testing. It implies that it can act adaptively, which means it reuses the policy wherever it is useful and does more exploration wherever required.

\subsubsection{Sensitivity Analysis}
\begin{figure}[h]
\includegraphics[width=0.98\textwidth, height=3.5cm]{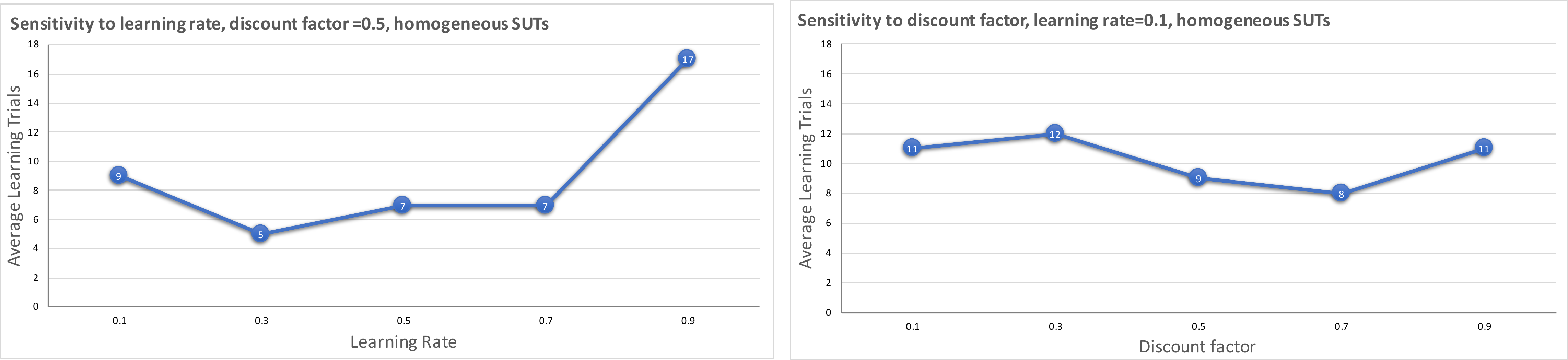}
\label{fig:Sensitivity of SaFReL-homogeneous set of SUTs-DiscountFact}
\begin{center}
\begin{tabular}{ |p{2.7cm}|p{1.3cm}|p{1.3cm}|p{1.3cm}|p{1.3cm}|p{1.3cm}|}
 \hline
 \multicolumn{6}{|c|}{\small Average Efficiency of SaFReL with $\epsilon=0.2$, Discount factor $\gamma=0.5$} \\
 & \small $\alpha=0.1$ &\small $\alpha=0.3$& \small $\alpha=0.5$ & \small $\alpha=0.7$ & \small $\alpha=0.9$ \\
 \hline
 \small Average number of learning trials & 9 &	5&	7&	7&	17\\
 \hline
 \hline
\end{tabular}
\begin{tabular}{ |p{2.7cm}|p{1.3cm}|p{1.3cm}|p{1.3cm}|p{1.3cm}|p{1.3cm}|}
 \hline
 \multicolumn{6}{|c|}{\small Average Efficiency of SaFReL with $\epsilon=0.2$, Learning rate $\alpha=0.1$} \\
 & \small $\gamma=0.1$ &\small $\gamma=0.3$& \small $\gamma=0.5$ & \small $\gamma=0.7$ & \small $\gamma=0.9$ \\
 \hline
 \small Average number of learning trials & 11 & 12 & 9 & 8 & 11\\
 \hline
\end{tabular}

\caption{Sensitivity of SaFReL to learning rate and discount factor on the homogeneous set of SUTs}
\label{fig:Sensitivity of SaFReL-homogeneous set of SUTs}
\end{center}
\end{figure}

\begin{figure}[h]
\includegraphics[width=0.98\textwidth, height=3.5cm]{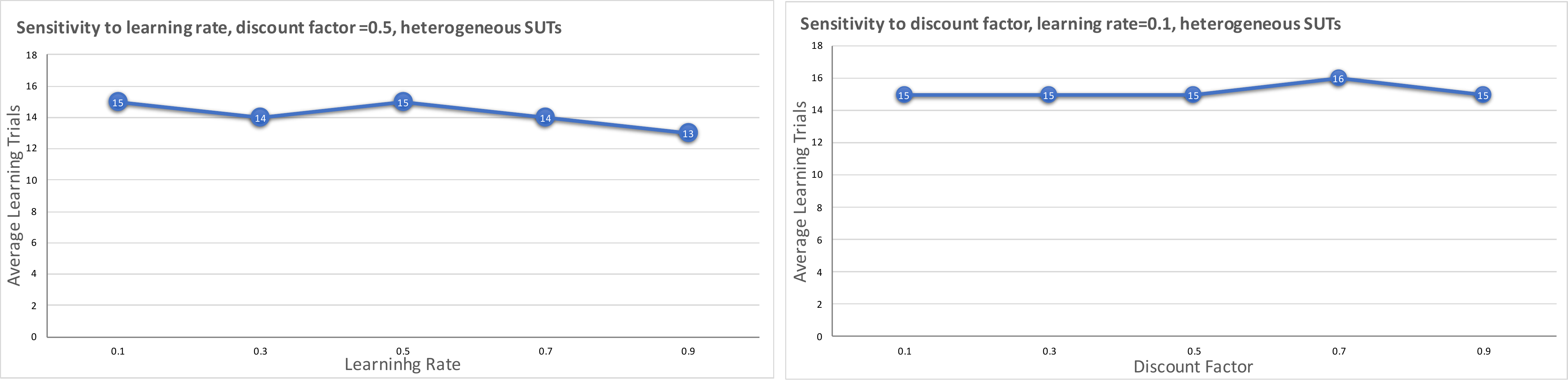}
\label{fig:Sensitivity of SaFReL-heterogeneous set of SUTs-DiscountFact}
\begin{center}
\begin{tabular}{ |p{2.7cm}|p{1.3cm}|p{1.3cm}|p{1.3cm}|p{1.3cm}|p{1.3cm}|}
 \hline
 \multicolumn{6}{|c|}{\small Average Efficiency of SaFReL with adaptive $\epsilon$, Discount factor $\gamma=0.5$} \\
 & \small $\alpha=0.1$ &\small $\alpha=0.3$& \small $\alpha=0.5$ & \small $\alpha=0.7$ & \small $\alpha=0.9$ \\
 \hline
 \small Average number of learning trials & 15&	14&	15&	14&	13\\
 \hline
 \hline
\end{tabular}
\begin{tabular}{ |p{2.7cm}|p{1.3cm}|p{1.3cm}|p{1.3cm}|p{1.3cm}|p{1.3cm}|}
 \hline
 \multicolumn{6}{|c|}{\small Average Efficiency of SaFReL with adaptive $\epsilon$, Learning rate $\alpha=0.1$} \\
 & \small $\gamma=0.1$ &\small $\gamma=0.3$& \small $\gamma=0.5$ & \small $\gamma=0.7$ & \small $\gamma=0.9$ \\
 \hline
 \small Average number of learning trials & 15&	15&	15&	16&	15\\
 \hline
\end{tabular}
\caption{Sensitivity of SaFReL to learning rate and discount factor on the heterogeneous set of SUTs}
\label{fig:Sensitivity of SaFReL-heterogeneous set of SUTs}
\end{center}
\end{figure}

To answer RQ3, we study the impacts of the learning parameters including learning rate ($\alpha$) and discount factor ($\gamma$), on the efficiency of SaFReL on both homogeneous and heterogeneous sets of SUTs. For conducting sensitivity analysis, we implement two sets of experiments that involve changing one learning parameter while keeping the other one constant. For the experiments running on a homogeneous set of SUTs, we use $\varepsilon$-greedy with $\varepsilon=0.2$ as the well-suited variant of action selection strategy with respect to the results of efficiency analysis (See Figure \ref{fig: Efficiency of SaFReL-homogeneous set of SUTs-transfer learning}) and on the heterogeneous set of SUTs, we use adaptive $\varepsilon$ selection (See Figure \ref{fig:Efficiency of SaFReL-heterogeneous set of SUTs-adaptive}). During the sensitivity analysis experiments, to study the impact of the learning rate changes, we set the discount factor to 0.5. While examining the impact of the discount factor changes, we keep the learning rate fixed to 0.1. Figure \ref{fig:Sensitivity of SaFReL-homogeneous set of SUTs} shows the sensitivity of SaFReL to changing learning rate and discount factor parameters when it acts on a homogeneous set of SUTs (CPU-intensive). Figure \ref{fig:Sensitivity of SaFReL-heterogeneous set of SUTs} depicts the results of the sensitivity analysis of SaFReL on a heterogeneous set of SUTs.

\section{Discussion} \label{sec::Discussion}
\subsection{\blue{Efficiency, Adaptivity and Sensitivity Analysis}} \textbf{RQ1:} Using multiple experiments, we studied the efficiency of SaFReL compared to a typical stress testing procedure, on both a set of homogeneous and heterogeneous SUTs regarding the use of different action selection strategies.
The results of the experiments running on a set of 50 CPU-intensive SUT instances as a homogeneous set of SUTs, Figure \ref{fig: Efficiency of SaFReL-homogeneous set of SUTs-transfer learning} and Tables \ref{table: Efficiency of SaFReL-homogeneous set of SUTs-transfer learning} and \ref{table: improvement-SaFReL-homogeneous set of SUTs-transfer learning}, show that using $\varepsilon$-greedy, $\varepsilon=0.2$ as action selection strategy in the transfer learning leads to desired efficiency and an improvement in the computation time (around $42\%$) compared to the typical stress testing. It causes SaFReL to rely more on reusing the learned policy and results in computation time saving. The existing similarity between the performance sensitivity of SUTs in a homogeneous set of SUTs makes the strategy of policy reusing successful in this type of testing situations.

Furthermore, we studied the efficiency of SaFReL on a heterogeneous set of 50 SUTs containing different CPU-intensive, memory-intensive and disk-intensive ones. The results of the analysis illustrate that choosing an action selection strategy without considering the heterogeneity among the SUTs (e.g., using the typical variants of $\varepsilon$-greedy) does not lead to desirable efficiency compared to the typical stress testing (See Figure \ref{fig:Efficiency of SaFReL-heterogeneous set of SUTs-typical epsilon}, Table \ref{table: Efficiency of SaFReL-heterogeneous set of SUTs-transfer learning} and \ref{table: improvement-SaFReL-heterogeneous set of SUTs-transfer learning}).
Then, we augmented our fuzzy RL-based approach with an adaptive action selection strategy that is a heterogeneity-aware strategy for adjusting the value of $\varepsilon$. It measures the similarity between the performance sensitivity of the SUTs and adjusts the $\varepsilon$ parameter. 
As shown in Figure \ref{fig:Efficiency of SaFReL-heterogeneous set of SUTs-adaptive}, using the adaptive $\varepsilon$-greedy, addressed the issue and led to an efficient generation of the target performance test case and a computation time improvement (around $31\%$). It makes the agent able to reuse the learned policy according to the conditions, which means it uses the learned policy wherever it is useful and does more exploration wherever it is required.

\textbf{RQ2:} At the last part of the efficiency and adaptivity analysis, we extended our analysis by measuring the adaptivity of SaFReL when it performs on a heterogeneous set of SUTs. As shown in Figure \ref{fig:Adaptivity of SaFReL-heterogeneous set of SUTs}, with the use of the adaptive $\varepsilon$-greedy, SaFReL is able to adapt to changing testing situations while preserving the efficiency. 

\textbf{RQ3:} The results of the sensitivity analysis experiments on the homogeneous set of SUTs show that adjusting the learning rate to lower values such as 0.1 leads to better efficiency. Furthermore, regarding the sensitivity analysis of SaFReL to the discount factor on a homogeneous set of SUTs, the experimental results depict that lower values of the discount factor are suitable choices for the desired operation that we expect.
However, the results of the sensitivity analysis on the heterogeneous set of SUTs do not show a considerable effects on the average efficiency of SaFReL when it acts on a heterogeneous set of SUTs regarding the use of adaptive $\varepsilon$-greedy. 
 
\subsection{Lessons Learned}
The experimental evaluation of SaFReL shows how machine learning can guide performance testing towards being automated and taking one step further towards being autonomous. Common approaches for generating performance test cases mostly rely on source code or system models, but such development artifacts might not always be available. Moreover, drawing a precise model of a complex system  predicting the state of the system upon given performance-related conditions requires a solid endeavor. This makes room for machine learning, particularly model-free learning techniques. Model-free RL is a machine learning technique enabling the learner to explore the environment (the behavior of the SUT on the execution platform in this case) and learn the optimal policy to accomplish the objective (finding the intended performance breaking point in this case) without having a model of the system. The learner stores the learned policy and is able to replay the learned policy in further suitable situations. This important characteristic of RL leads to a reduction in the effort of the learner to accomplish the objective in further cases and consequently leads to improved efficiency.  
\blue{Therefore, the main features that lead SaFRel to outperform an exploratory (search-based) technique are the capability of storing knowledge during the exploration and reusing the knowledge in suitable situations, and the possibility of selective and adaptive control on exploration and exploitation.}  

In general, automation, reduction of computation time and cost, and less dependency on source code and models are profound strengths of the proposed RL-assisted performance testing. 
Regarding applicability, according to the aforementioned strengths and the results of the experimental evaluation, the proposed approach could be beneficial to performance testing of software variants in software product lines, evolving software in continuous Integration/Delivery process and performance regression testing.   

\textit{Changes in Future Trends.} With the emergence of serverless architecture, which incorporates third-party backend services (BaaS) and/or runs the server-side logic in state-less containers that are fully-managed by providers (FaaS), a slight shift in the objectives of performance evaluation, particularly performance testing on cloud-native applications is expected. Within the serverless architecture, the backend code is run without the need to manage and provision the resources on servers. For example in FaaS, scaling, including the resource provisioning and allocation, is automatically done by the provider whenever it is needed, to preserve the response time requirement of the application. In general, regarding the capabilities of new execution platforms and deployment architectures, the objectives of performance testing might be slightly influenced. Nevertheless, it is still crucial for a wide range of software systems.

\subsection{Threats to Validity} Some of the main sources of threat to the validity of our experimental evaluation results are as follows:

\blue{\textit{Construct.} One of the main sources of threat is the formulation of the RL technique to address the problem, which is very important for successful learning. Modeling the state space, actions, and also the reward function are major players to guide the agent throughout the learning and make it learn the optimal policy. For example, boundaries defined in discrete states modeling are a threat to internal validity. To mitigate this threat, we used a fuzzy labeling technique to deal with the issue of uncertainty in defining sharp values for boundaries. Regarding the actions, the formulation of actions affects the granularity of the exploration steps, thus we tried to define actions in a way to provide reasonable granularity for the exploration steps.}

\textit{Internal.} There are a number of threats to the internal validity of the results. RL techniques like many other machine learning algorithms are influenced by their hyperparameters such as learning rate and discount factor. During our efficiency and adaptivity analysis experiments, we did not change the learning parameters, we also conducted a set of controlled experiments to study the influence of learning parameters on the efficiency of our approach.

\blue{The insufficient number of learning episodes/iterations could also act as a source of threat in the initial learning. To alleviate this threat, we iterated the initial learning sufficiently to ensure the convergence.} 
\blue{Moreover, using a performance simulation module instead of executing SUTs actually is considered as a source of threat to the validity of results.}  

\blue{Finally, model-free RL is mainly intended to solve a decision-making problem (to find an optimal policy to behave) without access to a model of the environment. Therefore, not considering the structure of the environment might be a source of threat in case of improper formulation of the RL technique.}

\textit{External.} Model-free RL learns the optimal policy to achieve the target through interaction with the environment. Our approach was formulated based on the SUTs with three types of performance sensitivity that are CPU-intensive, memory-intensive, and disk-intensive, and our results are derived from the experimental evaluation of our approach on these types of SUTs. If the experiment contains SUTs with other types of performance sensitivity such as network-intensive programs, then the approach needs to be reformulated slightly to support new types of performance sensitivities.

\blue{Moreover, the dependency of the performance simulation module on the performance sensitivity values could raise a threat to validity in case of deploying the smart tester agent with the performance simulation module. The performance simulation module requires the performance sensitivity values for the SUTs as we described in our experiments. However, given a real deployment of the approach, e.g., in a cloud-based testing setup without the performance simulation module, the dependency on the performance sensitivity values are lighter and their exact values are not necessary. Nonetheless, it is still considered as a source of threat.}    

\section{Related Work} \label{sec::Related Work}
Measurement of performance metrics under typical or stress test execution conditions, which involve both workload and platform configuration aspects \cite{ menasce2002load, hill2009tools, apte2017autoperf, michael2017cloudperf, jindal2019performance}, detection of performance-related issues such as functional problems or violations of performance requirements emerging under certain workload or resource configuration conditions \cite{briand2005stress, zhang2011automatic, ayala2018one, schulz2019behavior} are common objectives of different types of performance testing.

Different approaches have been proposed to design the target performance test cases for accomplishing performance-related objectives such as finding intended performance breaking points. Performance test conditions involve both workload and resource configuration status. A general high-level categorization of main techniques for generating the performance test cases is as follows:

\textit{Source code analysis.} Deriving workload-based performance test conditions using data-flow analysis and symbolic execution are examples of techniques for designing fault-inducing performance test cases based on source code analysis to detect performance-related issues such as functional problems (like memory leaks) and performance requirement violations \cite{yang1996towards, zhang2011automatic}.

\textit{System model analysis.} Modeling the system behavior in terms of performance models like Petri nets and using constraint solving techniques \cite{zhang2002automated}, using the control flow graph of the system and applying search-based techniques \cite{gu2009search, di2007search}, and using other types of system models like UML models and using genetic algorithms \cite{garousi2010genetic, garousi2008empirical, garousi2008traffic, costa2012generating, da2011generation} to generate the performance test cases are examples of the techniques based on system model analysis for generating performance test cases.

\textit{Behavior-driven declarative techniques.} Using a Domain Specific Language (DSL) to provide declarative goal-oriented specifications of performance tests and model-driven execution frameworks for automated execution of the tests \cite{ferme2018declarative, ferme2017towards, walter2016asking}, and using a high-level behavior-driven language inspired from Behavior-Driven Development (BDD) techniques to define test conditions \cite{schulz2019behavior} in combination with a declarative performance testing framework like BenchFlow \cite{ferme2017towards} are examples of behavior-driven techniques for performance testing.

\textit{Modeling the realistic conditions.} Modeling the real user behavior through stochastic form-oriented models \cite{draheim2006realistic, lutteroth2008modeling}, extracting workload characteristics from the recorded requests and modeling the user behavior using, e.g., extended finite state machines (EFSMs) \cite{shams2006model} or Markov chains \cite{vogele2018wessbas}, sandboxing services and deriving a regression model of the deployment environment based on the data resulting from sandboxing to estimate the service capacity \cite{jindal2019performance}, end-user clustering based on the business-level attributes extracted from usage data \cite{maddodi2018generating}, and using automated GUI testing tools with capture and replay techniques to generate realistic interactive usage sequences \cite{adamoli2011automated} are examples of techniques based on modeling the realistic conditions to generate the performance test cases.

\textit{Machine learning-enabled techniques.} Machine learning techniques such as supervised and unsupervised algorithms mainly work based on building models and extracting patterns (knowledge) from the data. While, some other techniques such as RL algorithms are intended to train the learner agent to solve the problems (tasks). The agent learns an optimal way to achieve an objective through interacting with the system. Machine learning has been widely used for analysis of data resulting from the performance testing and also for performance preservation. For example, anomaly detection through analysis of performance data, e.g., resource usage, using clustering techniques \cite{syer2011identifying}, predicting reliability from the testing data using Bayesian Networks \cite{avritzer2008reliability}, performance signature identification based on performance data analysis using supervised and unsupervised learning techniques \cite{malik2013automatic, malik2010automatic}, and also adaptive RL-driven performance in particular response time control for cloud services \cite{ibidunmoye2017adaptive, veni2016auto, jamshidi2016fuzzy} and also software on other execution platforms, e.g., PLC-based real-time systems \cite{moghadam2018adaptive}.
Machine learning has been also applied to the generation of performance test cases in some studies. For example, using symbolic execution in combination with an RL algorithm to find the worst-case execution path within a SUT \cite{koo2019pyse}, using RL to find a sequence of input workload leading to performance degradation \cite{ahmad2019exploratory}, and a feedback-driven learning to identify the performance bottlenecks through extracting rules from execution traces \cite{grechanik2012automatically}. There are also some adaptive techniques slightly analogous to the concept of RL for generating performance test cases. For example, an adaptive workload generation that adapts the workload dynamically based on some pre-defined adjustment policies \cite{ayala2018one}, and a feedback-driven approach that uses search algorithms to benchmark an NFS server based on varying workload parameters to find the workload peak rate reaching the target response time confidence level.


\section{Conclusion} \label{sec::Conclusion}

Performance testing is a family of techniques commonly used as part of performance analysis, e.g., estimating performance metrics or detecting performance violations. 
One important goal of performance testing, particularly in mission-critical domains, is to verify the robustness of the SUT in terms of finding performance breaking point. Model-driven techniques might be used for this purpose in some cases, but drawing a precise model of the performance behavior of a complex software system under different application-, platform- and workload-based affecting factors is difficult. Furthermore, such modeling might disregard important implementation and deployment details. In software testing, source code analysis, system model analysis, use-case based design, and behavior-driven techniques are some common approaches for generating performance test cases. However, source code or other artifacts might not be available during the testing.

In this paper, we proposed a fuzzy reinforcement learning-based performance testing framework (SaFReL) that adaptively and efficiently generates the target performance test cases resulting in the intended performance breaking points for different software programs, without access to source code and system models. We used Q-learning augmented by fuzzy state modeling and an action selection strategy adaptation that resulted in a self-adaptive autonomous tester agent. The agent can learn the optimal policy to achieve the target (reaching the intended performance breaking point), reuse its learned policy when deployed to test similar software and adapt its strategy when targeting software with different characteristics.

We evaluated the efficiency and adaptivity of SaFReL through a set of experiments based on simulating the performance behavior of various SUT programs. During the experimental evaluation, we tried to answer how efficiently and adaptively SaFReL can perform testing of different SUT programs compared to a typical stress testing approach. We also performed a sensitivity analysis to explore how the efficiency of SaFReL is affected by changing the learning parameters. 

We believe that the main strengths of using the intelligent automation offered by SaFReL are 1) efficient generation of test cases and reduction of computation time, and 2) less dependency on source code and models. 
Regarding applicability, we believe that SaFReL could be beneficial to the testing of software variants, evolving software during the (CI/CD) process, and regression performance testing. Applying some heuristics and techniques to speed up the exploration of the state space like using multiple cooperating agents, and also extending the proposed approach to support workload-based performance test cases are further steps to continue this research.

\section*{Acknowledgment}
This work has been supported by and received funding partially from the TESTOMAT, XIVT, IVVES and MegaM@Rt2 European projects.

\bibliography{main}
\bibliographystyle{unsrt}

\end{document}